\documentclass[preprint,a4paper,showabstract,superscriptaddress]{revtex4-2}

\usepackage{times,ulem,color,bm}
\usepackage{amsfonts}

\usepackage{dutchcal}

\newcommand{\review}[1]{\textcolor{black}{#1}}
\newcommand{\rev}[1]{\textcolor{black}{#1}}
\newcommand{\revrita}[1]{\textcolor{black}{#1}}

\newcommand{\alex}[1]{\textcolor{black}{#1}}

\usepackage{xcolor, soul}
\sethlcolor{pink}

\usepackage{hyperref,graphicx}    

\usepackage{natbib}
\usepackage{graphicx}
\usepackage{amsmath,amsfonts,amssymb}
\usepackage[inline]{enumitem}
\usepackage{textcomp} 
\usepackage{gensymb} 
\usepackage[mathscr]{euscript}

\usepackage{booktabs}
\usepackage{siunitx}
\usepackage{url}
\usepackage{esint}
\usepackage{xcolor}
\usepackage[utf8]{inputenc}

\usepackage{titlesec}

\usepackage{setspace} 

\usepackage{xr} 

\usepackage{ragged2e}

\def\rout{\mathbf{r}_\textrm{out}}
\def\rin{\mathbf{r}_\textrm{in}}

\def\kout{\mathbf{k_\mathrm{out}}}

\def\kpl{\mathbf{k_\mathrm{+}}}
\def\kpr{\mathbf{k^\mathrm{'}}}
\def\kin{\mathbf{k_\mathrm{in}}}

\def\Rrr{\mathbf{R}_{\bm{\rho\rho}}}

\def\Rkk{\mathbf{R}_{\mathbf{kk}}}
\def\Rkkp{\mathbf{R}'_{\mathbf{kk}}}

\def\Rrrb{\overline{\mathbf{R}}_{\bm{\rho\rho}}}

\def\r{\mathbf{r}}
\def\k{\mathbf{k}}
\def\s{\mathbf{s}}
\def\rp{\mathbf{r}_\textrm{p}}

\begin{document}

\title{Imaging the crustal and upper mantle structure of the North Anatolian Fault: A Transmission Matrix Framework for Local Adaptive Focusing}

\author{Rita Touma}
\affiliation{Institut Langevin, ESPCI Paris, PSL University, CNRS, Paris, France}
\affiliation{ISTerre, Universit\'{e} Grenoble Alpes, Maison des G\'{e}osciences, BP 53, F-38041 Grenoble, France}
\author{Arthur Le Ber}
\affiliation{Institut Langevin, ESPCI Paris, PSL University, CNRS, Paris, France}
\author{Michel Campillo}
\affiliation{ISTerre, Universit\'{e} Grenoble Alpes, Maison des G\'{e}osciences, BP 53, F-38041 Grenoble, France}
\author{Alexandre Aubry}
\affiliation{Institut Langevin, ESPCI Paris, PSL University, CNRS, Paris, France}
\email{alexandre.aubry@espci.fr}

\date{\today}
\begin{abstract}
Imaging the structure of major fault zones is essential for our understanding of crustal deformations and their implications on seismic hazards. Investigating such complex regions presents several issues, including the variation of seismic velocity due to the diversity of geological units and the cumulative damage caused by earthquakes. Conventional migration techniques are in general strongly sensitive to the available velocity model. Here we apply a passive matrix imaging approach which is robust to the mismatch between this model and the real seismic velocity distribution. This method relies on the cross-correlation of ambient noise recorded by a geophone array. The resulting set of impulse responses form a reflection matrix that contains all the information about the \revrita{subsurface}. In particular, the reflected body waves can be leveraged to: (i) determine the transmission matrix between the Earth's surface and any point in the subsurface; (ii) build a confocal image of the subsurface reflectivity with a transverse resolution only limited by diffraction. As a study case, we consider seismic noise (0.1-0.5 Hz) recorded by the Dense Array for Northern Anatolia (DANA) that consists of 73 stations deployed for 18 months in the region of the 1999 Izmit earthquake. Passive matrix imaging reveals the scattering structure of the crust and upper mantle around the NAFZ over a depth range of 60 km. The results show that most of the scattering is associated with the Northern branch that passes throughout the crust and penetrates into the upper mantle. 
\end{abstract}
\maketitle
\section{Introduction}
The North Anatolian Fault zone (NAFZ) is one of the major continental right-lateral strike slip faults, and forms a border between the Eurasian continent and the Anatolian block. With an extremely well developed surface expression, it is one of the most active faults in the Eastern Mediterranean region~\citep{sengor1979,barka1992}. It is over $1600 $ km long and extends from eastern Turkey in the east to Greece in the west and, historically, has been subject to many destructive earthquakes~\citep{ambraseys1995,stein1997}. The seismic activity of such large faults constitutes a continuous hazard/threat to the surrounding regions and big cities, especially Istanbul city located to the West of the fault. 

Faults are well defined at the surface by the localized deformation and displacement delineating the fault traces, but \revrita{their} deep structure remains poorly understood~\citep{vauchez2012}. The understanding of such major fault systems and seismic hazard requires a characterization of the geometrical and seismic properties of the crust and upper mantle. 
A large number of geological and geophysical studies \revrita{have} discussed the complexity of fault zones {and their relation with their deep roots~\citep{stein1997}}. They are not only confined in the mid crust\rev{;} {indeed, models suggest that they} penetrate deep into the crust and extend to the upper Mantle~\citep{lyakhovsky2009}. If so, faults develop into shear zones, {corresponding to} a volume of localized deformation {accounting for the relative displacement of the tectonic blocks. } 

Seismic imaging techniques, especially reflection, refraction, and tomographic methods, constitute a very powerful tool to characterize fault zones and report the variation of the properties of the crust and the upper mantle. They rely on the study of wave propagation inside the Earth that is governed by the density and elastic properties of the rocks. 
To properly probe the medium, waves should be generated by a dense distribution of seismic sources. Conventional seismic exploration techniques use either earthquakes as seismic \revrita{sources}, or explosions and vibrators to generate seismic waves in regions with weak seismicity. Because of the limitations {in the earthquake distribution }and high cost of active methods, there \revrita{is} a need for alternative imaging approaches that would not rely on any coherent source. In the 2000's, the extraction of deterministic information about the Earth structure from ambient seismic noise revolutionized the field of seismology~\cite[see e.g.][]{campilloandroux2014}. It was shown that the cross-correlation of {diffuse waves or} ambient seismic noise recorded at two stations provides \review{an \revrita{}{estimate} of }the Green's function between those two stations~\cite[see e.g.][]{campillo}. The reflection response of the medium is then retrieved and can be applied to build tomographic or structural images of the Earth. Because ambient noise is {dominated} by surface waves, their Green's function component can be easily extracted~\citep{shapiro2004}. It has been \revrita{demonstrated} that\rev{, under energy equipartition,} body wave reflections can also be retrieved from ambient seismic noise cross-correlations~\citep{draganov2007,poli2012a,poli2012b}. Body waves contain valuable information on the structure of the medium {in depth} and can be investigated to obtain high-resolution images of the crust and the mantle~\citep{Retailleau2020}. 

Faults are usually imaged indirectly through strong velocity contrasts in tomographic profiles~\citep{zigone2019imaging}, or through the offset of geological layers observed in reflectivity images~\citep{qian2020}. {However, tomographic images exhibit a relatively \revrita{poor} resolution, while reflection imaging methods are strongly sensitive to the available velocity model. Interestingly, a reflection matrix approach has been recently proposed to cope with these issues. Originally developed in acoustics~\citep{lambert2020,Lambert2020b} and optics~\citep{kang2017,Badon2019}, this approach has been recently applied to passive seismology~\citep{blondel2018,touma2021}. By considering high frequency seismic noise ($10$-$20$ Hz), high resolution images of complex areas, such as volcanoes~\citep{giraudat2021} and fault zones~\citep{touma2021}, have been obtained over a few km depth. In this paper, we aim to characterize the crustal structure of the NAFZ  at a much larger scale (until $60$ km depth). To that aim, a lower frequency bandwidth ($0.1$-$0.5$ Hz) has been considered. At the corresponding wavelengths, the subsurface reflectivity can be considered as continuous rather than being seen as a discrete distribution of scatterers as in our previous works~\citep{giraudat2021,touma2021}. As we will see, this \rev{continuous}  reflectivity can be exploited to enable a local and adapted auto-focus on each part of the subsurface image, thereby showing {an important} robustness to the inaccuracy of the initial wave velocity model.}


{Seismic matrix imaging} is based on the passive measurement of the reflection matrix $\mathbf{R}$ associated with a network of geophones. It contains the set of impulse responses between each \review{pair of} geophones extracted from cross-correlations of seismic noise.  
Based on the available velocity model, a focused reflection (FR) matrix is built by applying a redatuming process to $\mathbf{R}$~\citep{blondel2018,lambert2020}. \revrita{This matrix} contains the impulse response between virtual sources and receivers synthesized inside the medium. \rev{In the following, matrix ``input'' and ``output'' will refer to virtual sources (downgoing waves) and receivers (upgoing waves), respectively.}
\rev{This FR matrix is powerful as it first provides an image of the subsurface reflectivity by considering its diagonal elements \rev{(\textit{i.e} when virtual source and receiver coincide); this is the \textit{so-called} confocal image}. Moreover, its off-diagonal elements allow a local quantification of aberrations in the vicinity of each virtual source. Those aberrations correspond to the imperfections of the image induced by the mismatch between the wave velocity model and the actual seismic velocity distribution in the \revrita{subsurface}.} {In contrast with previous works~\citep{giraudat2021,touma2021}, a multi-layered wave velocity model is here considered rather than just an homogeneous model. This more sophisticated description of seismic wave propagation enables a better time-to-depth conversion in the confocal image and a better focusing process. Nevertheless, the FR matrix still highlights residual aberrations that result from} the mismatch between the velocity model and the actual velocity distribution. The fluctuations of wave velocities actually induce phase distortions on the focused wave-fronts that result in a blurry image {of the NAFZ subsurface.}

{To overcome these detrimental effects, the FR matrix can be first projected in a plane wave basis. By exploiting the angular input-output correlations of the reflection matrix, phase distortions of the incident and reflected wave-fronts can be identified and compensated. This is the principle of the CLASS algorithm (acronym for closed-loop accumulation of single scattering), originally developed in optical microscopy~\citep{kang2017, choi2018, yoon2019}. Applied for the first time to seismology in the present study, CLASS successfully compensates for spatially-invariant aberrations and will be shown to clearly improve the confocal image of the NAFZ subsurface.}

{Nevertheless, high-order aberrations subsist and are addressed through the distortion matrix concept in a second step. Originally introduced in ultrasound imaging~\citep{Lambert2020b} and optical microscopy~\citep{Badon2019}, this operator contains the phase distortions of the incident and reflected wave-fronts with respect to the propagation model. It was recently exploited in passive seismology in order to image the San Jacinto Fault zone scattering structure that exhibits a sparse distribution of scatterers~\citep{touma2021}. Here, we apply it in a new scattering regime since the NAFZ subsurface exhibits, in the frequency range under study, a continuous reflectivity distribution made of specular reflectors and randomly distributed heterogeneities. In this regime, a local time reversal analysis of the distortion matrix can be performed in order to retrieve the transmission matrix between the Earth's surface and any point of the subsurface~\citep{lambert2021distor}. This transmission matrix is a key tool since
its phase conjugate provides the optimized focusing laws that need to be applied to the reflection matrix in order to retrieve a diffraction-limited image of the subsurface. While most conventional reflection imaging techniques are strongly sensitive to the available velocity model, the reflection matrix approach is robust with respect to its \revrita{limitations}. An approximate velocity distribution is actually sufficient since a time-reversal analysis of seismic data enables a local and adapted
auto-focus on each part of the subsurface image.}

To image the crustal structure of NAFZ, we use data from the Dense Array for Northern Anatolia~\citep{DANA2012} that was deployed over the western segment of the fault, in the latest rupture region during the 1999 Izmit ($M = 7.6$) and D\"uzce ($M = 7.2$) earthquakes~\citep{barka2002,akyuz2002}. The dense array was installed temporarily between May 2012 and October 2013. It consists of 73 3-component broadband seismometers, 66 stations arranged along $11$ east-west lines and 6 North-South lines forming a rectangular grid and covering an area of $35$ km by $70$ km with a nominal inter-station spacing of $\sim$7km (Fig.~\ref{chap3fig1}a). Seven additional stations were deployed east of the rectangular array in a semi-circle shape.
In this region, the fault splits into two major strands: the northern (NNAFZ) and southern (SNAFZ) strands (Fig.~\ref{chap3fig1}b). The northern strand, where most of the {continuous deformation occurs according to geodetic studies}~\citep{barka1992,reilinger2006}, has been subject to a series of major earthquakes in the last century, {among them the 1999 Izmit Earthquake.} On the contrary, the latest rupture of the southern branch dates back to the fifteenth century~\citep{ambraseys2002}. The fault delineates three tectonic blocks (Fig.~\ref{chap3fig1}b): $\textit{(i)}$ the Istanbul Zone (IZ) situated North of the northern branch, $\textit{(ii)}$ the Sakarya zone (SZ) situated to the South of the southern branch and $\textit{(iii)}$ the Armutlu-Almacik crustal block (AA) located in the center, between the two fault strands{~\citep{yilmaz1995,okay1999,chen2002}}.
{Differences in crustal composition and properties between these blocks have been reported. Strong velocity contrasts were found across the fault strands by several tomographic studies~\citep{salah2007,koulakov2010,papaleo2017,papaleo2018} and full waveform inversion studies~\citep{fichtner2013,ccubuk2017}. 
Low velocity zones are found below the surface traces of the SNAFZ and NNAFZ~\citep{papaleo2017,papaleo2018}. 
The crust of Istanbul and Armutlu Blocks is characterized by high velocities while SZ shows relatively low velocities~\citep{koulakov2010,papaleo2017,papaleo2018,taylor2019}.}

The present study reveals the 3D scattering structure of the medium below this major fault. It does not only image planar interfaces, but provides a direct insight on the heterogeneities that mainly sit in the vicinity of the strands. 
The observed results complement previous studies conducted in the region. A step in the Moho is detected below the Northern branch, and several sub-Moho structures are observed in the North confirming that the northern branch penetrates in the upper mantle. The southern strand does not have a strong signature in the scattering profiles. 

\section{{Passive Seismic Matrix Imaging}}

\subsection{Reflection matrix in the geophones basis}
\begin{figure*}
 \centering
 \includegraphics[width=\textwidth]{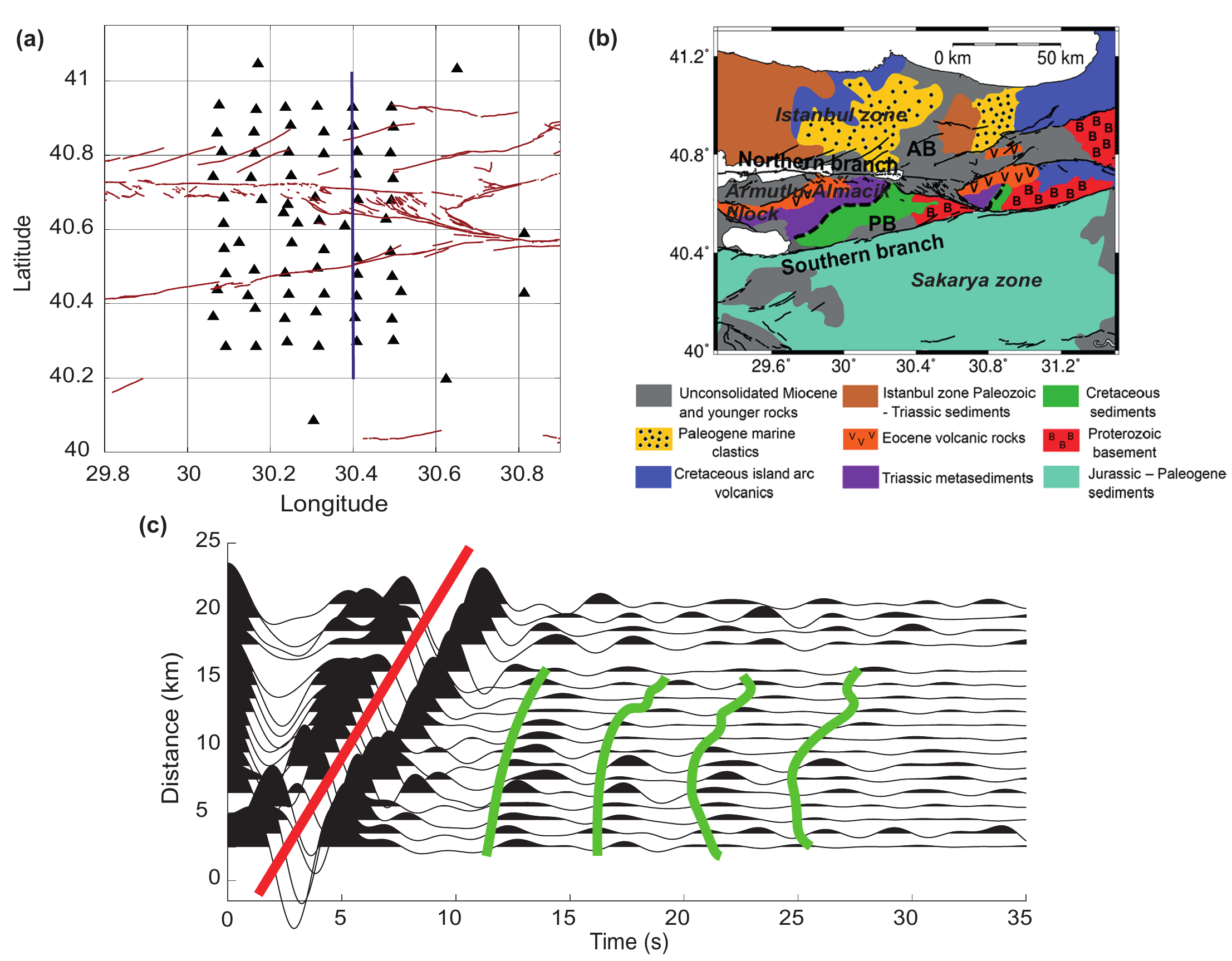}
 \caption{Study region and location of DANA array~\citep{DANA2012}. (a) Map of the study region and location of DANA array geophones (black triangle. The surface traces of the NAFZ are represented by the red lines~\citep{emre2018}. The blue line indicate the location of the cross sections represented in Figs.~\ref{chap3fig5}a and~\ref{chap3fig5_2}a. (b) Geological map of the region (adjusted from~\cite{taylor2019} and~\cite{akbayram2016}). 
 The major geological blocks are represented: Istanbul zone (IZ) in the North, Armutlu-Almacik (AA) block in the center and Sakarya zone (SZ) in the South. The Adapazari and Pamukova basin location are indicated by AB and PB, respectively.
 \review{(c) Ambient noise EE cross-correlation filtered between $0.1$ and $0.5$ Hz. The correlations between pairs of stations located South of the SNAF and having at least an angle of $45\degree$ with the East-West direction are plotted. These cross-correlograms are stacked over different distances. \rev{The predominant surface wave contribution and shear wave echoes induced by planar reflectors in the subsurface are highlighted by red and green lines, respectively.} }}
\label{chap3fig1}
  \end{figure*}

To apply {matrix imaging}, we used the ambient seismic noise recorded at DANA (see Fig.~\ref{chap3fig1}) to compute the cross-correlation functions of horizontal EE component over the 18 months of recording period. \rev{The choice of the EE component is made because it displays a better signal-to-noise than the NN component.}
{\revrita{With this choice of} body wave component, the waves being dealt with are \rev{mostly} shear waves that have been reflected \rev{in depth}.} First, the data were down-sampled at $25$ Hz and corrected from instrument response. Then, the data were split into one-hour windows. Each window is band-pass filtered between $0.1$ and $0.5$ Hz after applying a spectral whitening between $0.01$ and $1$ Hz~\citep{bensen2007}. The cross-correlation between each pair of stations is computed over one-hour windows and finally stacked to obtain the mean cross-correlation function \rev{with time lags ranging from $-35$ to $+35$ s}. \rev{The causal and the anti-causal parts of the cross-correlations are then summed in order to improve the convergence towards the Green's functions between seismic stations.}
Although, considering \rev{seismic} noise in {a} higher frequency range would allow, in principle, to improve the resolution of the images, matrix imaging requires the Nyquist criterion to be fulfilled: The inter-station distance ($7$ km) shall be of the order of a half-wavelength. Considering a S-wave velocity $c_0=1700$ m/s near the surface, this criterion led us to choose  the $0.1-0.5$ Hz frequency range ($\lambda = 5.7$ km at the central frequency\rev{, with $\lambda$ the wavelength at the Earth surface}). The ambient noise energy in the frequency band considered in this study comes from the secondary microseisms ($5-10$ s period band) \revrita{produced in} the ocean~\citep{longuet1950,hasselmann1963,stehly2006} and constitutes one of the most energetic parts of the seismic noise.

The symmetric cross-correlations can be stacked in a time-dependent response matrix {$\mathbf{R}_{\mathbf{ss}}(t)$}. 
One element \rev{$R(\mathbf{s}_i,\mathbf{s}_j,t)$} of this matrix corresponds to the impulse response between geophones located at positions \rev{$\mathbf{s}_i$ and $\mathbf{s}_j$}. In other words, \rev{$R(\mathbf{s}_i,\mathbf{s}_j,t)$} contains the seismic wave-field recorded at receiver \rev{$\mathbf{s}_i$} if a pulse was emitted by the virtual source \rev{$\mathbf{s}_j$} at time $t=0$. \rev{In the following, we will thus refer to columns and lines of the response  matrix as input and output wave-fields, respectively.} 

\rev{Figure~\ref{chap3fig1}c shows the impulse responses between pairs of stations located South of the SNAF and having at least an angle of $45\degree$ with the East-West direction because of the EE polarization considered in this study. These responses are stacked for different inter-station distances. Figure~\ref{chap3fig1}c is dominated by surface-wave energy travelling at about $3000$ m/s. Given the horizontal polarization of seismic waves considered here, these surface waves correspond to Love waves. Interestingly, some nearly vertical weaker events are also observed at larger times. Due to their horizontal polarization, they correspond to shear waves reflected by the \review{subsurface} heterogeneities in depth.}
  
\begin{figure*}
 \centering
    \includegraphics[width=\textwidth]{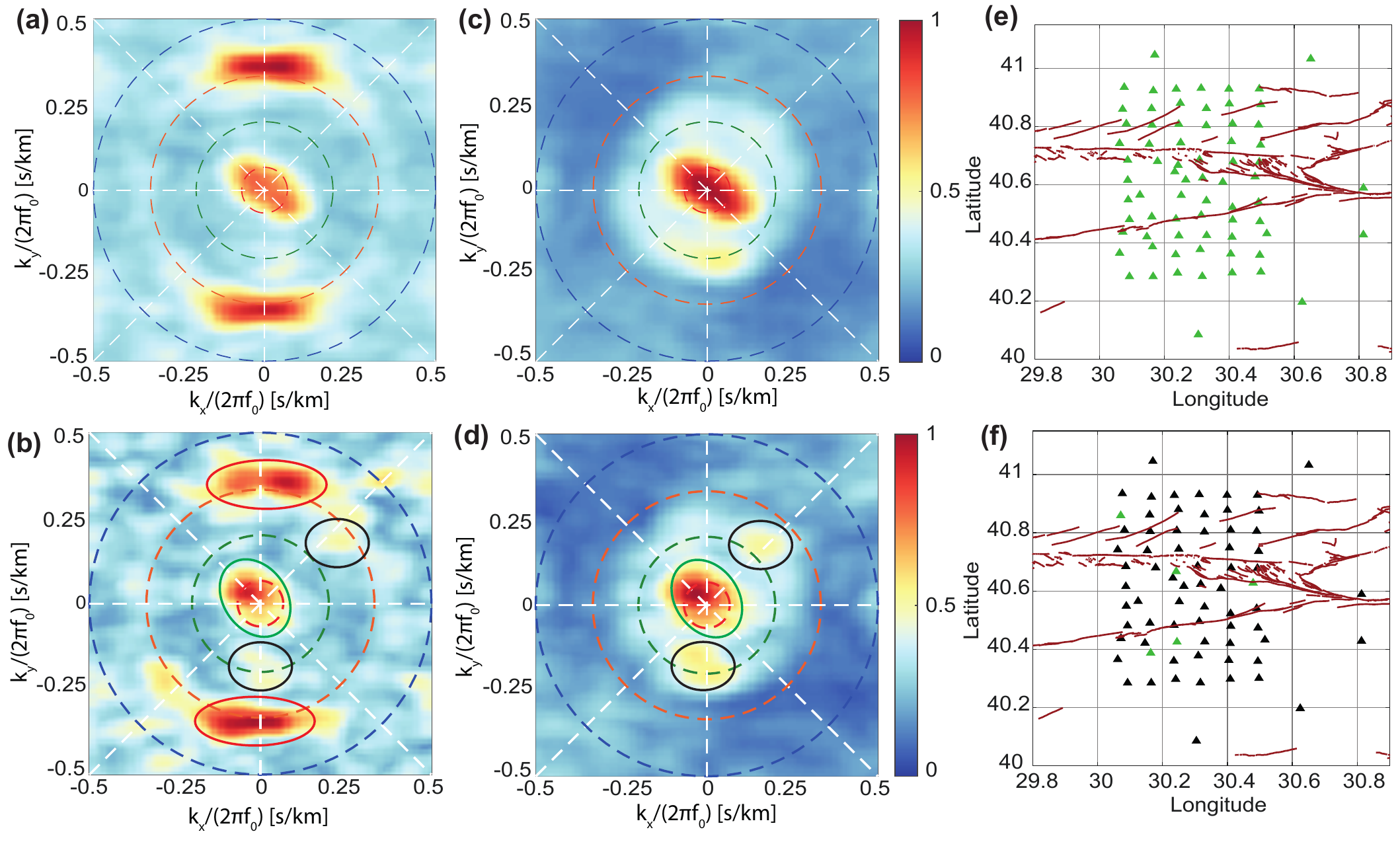}
 \caption{\rev{Apparent slowness of seismic echoes contained in the response matrix. (a,b) Plane wave decomposition of the Earth's response matrix $\mathbf{R_{ss}}$ at output ($f_0=0.2$ Hz, Eq.~\ref{Rsk_0}) averaged over the set of green geophones displayed in panels (e,f), respectively.  (c,d) Plane wave decomposition of the filtered response matrix $\mathbf{R'_{ss}}$ at output ($f_0=0.2$ Hz, z=25 km, Eq.~\ref{K_prim}) averaged over the set of green geophones in panels (e,f), respectively. The dashed circles correspond to apparent velocities of $5000$ m/s (green), $3000$ m/s (red) and $2000$ m/s (blue). Main echoes associated with Love waves (red ellipse), vertical shear waves (green ellipse) and off-axis shear waves (blue circle) are also highlighted in panels (b,d).} }
\label{fig10}
  \end{figure*}

\rev{Another way to highlight the surface and bulk wave components is to investigate the response matrix $\mathbf{R_{ss}}$ in the frequency domain. To \revrita{this} aim, a temporal Fourier transform is applied to $\mathbf{R_{ss}}(t)$. For each frequency $f$ in the bandwidth of interest (0.1-0.5 Hz), a monochromatic matrix $\overline{\mathbf{R}}_\mathbf{ss}(f)$ is obtained. The different wave components can then be discriminated by a plane wave decomposition of the output wave-fields, such that
\begin{equation}
\label{Rsk_0}
\review{\overline{\mathbf{R}}_\mathbf{sk}(f)= \overline{\mathbf{R}}_\mathbf{ss} (f) \times \mathbf{P}_{0},}
\end{equation}
\revrita{where the symbol $\times$ stands for the standard matrix product}. $\mathbf{P}_0=[P_0({\mathbf{s},\mathbf{k}})]$ is the Fourier transform operator that connects each geophone's position $\mathbf{s}$ to the transverse wave vector $\mathbf{k}=(k_x,k_y)$ of each angular component of the wave-field:
 \begin{equation}
{P}_0(\mathbf{s},\mathbf{k}) = \exp{\left(i {\mathbf{k}}\cdot \s \right), }
\label{Gsk}
\end{equation}
where the symbol $\cdot $ denotes the scalar product. Figure~\ref{fig10}a shows the result of this plane wave decomposition by displaying the mean angular distribution of the output wave-field at $f_0=0.2$ Hz. More precisely, this distribution is displayed as a function of the ratio between spatial frequencies $k_x/(2\pi)$ and $k_y/(2\pi)$ and frequency $f$. In that representation, surface waves emerge along a circle whose radius should correspond to the slowness $c_L^{-1}$ of Love waves, with $c_L\sim 3000$ m.$s^{-1}$ \citep{taylor2019}. On the \revrita{other hand}, reflected bulk waves are distributed over a disk of radius $c_0^{-1}$.} 

\rev{Figure~\ref{fig10}a clearly reveals: (\textit{i}) The contribution of Love waves with a \revrita{dominant intensity} lying along the y-direction; (\textit{ii}) A bulk wave component arising at small spatial frequencies and corresponding to the nearly vertical echoes already highlighted by Fig.~\ref{chap3fig1}c. The latter contribution is \textit{a priori} induced by extended reflectors in depth. On the contrary, diffuse scattering can generate reflected waves over a larger distribution of angles. It can therefore account for the incoherent background observed in Fig.~\ref{fig10}a. This background could result from the averaging of the random speckle pattern exhibited by the angular distribution of the reflected wave-field when only a few sources are considered  (Fig.~\ref{fig10}b). Nevertheless, it is difficult at this stage to be more affirmative since an imperfect convergence of noise correlations could also lead to such a random wave-field.}
  
\rev{A redatuming process is thus required to enhance the weight of scattered shear waves from seismic noise correlations and image the reflectivity of the deep structures around the NAFZ. }



\subsection{{Redatuming process}}
\label{redat}

An image of the medium reflectivity can be obtained by applying a double focusing operation to $\mathbf{R}_{\mathbf{ss}}$~\citep{blondel2018,touma2021}. It consists in back-propagating the response measured at the surface into wave-fields below the surface as if there were sources and receivers inside the medium. This is similar to the "wave-field extrapolation" concept that forms the basis of the migration process~\citep{berkhout1981}. It requires performing beamforming operations both at emission and reception. On the one hand, focusing in emission consists in applying appropriate time delays to the emitted sources so that waves constructively interfere and focus on one point inside the medium. Physically, this operation amounts to \revrita{synthesizing} a virtual source inside the medium. On the other hand, focusing in reception is carried out by applying proper time delays to the received signals so that they can constructively interfere. As in emission, this focusing operation can be seen as the synthesis of a virtual receiver inside the medium. This operation is known as "redatuming" in seismology~\citep{BerkhoutandWapenaar1993} and consists of virtually moving sources and receivers from the surface to the medium below (Fig.~\ref{chap3fig2}a). Generally, an image of the sub-surface is built by considering the response of virtual source and receiver placed at the same location (Fig.~\ref{chap3fig2}{f}). On the \revrita{other hand}, the principle of matrix imaging {consists in decoupling} both locations~\citep{lambert2020}.

{In the following, the reflection matrix will be expressed in three different bases: $\textit{(i)}$ the geophones basis where the matrix $\mathbf{R_{ss}}$ represents the cross-correlations between all pairs of stations located at $\mathbf{s}(x_s,y_s,0)$, $\textit{(ii)}$ the focused basis corresponding to the location $\mathbf{r}(x,y,z)$ of virtual sources and receivers synthesized by the focusing operations and in which the image of the medium is built, and $\textit{(iii)}$ the spatial Fourier basis \rev{$\mathbf{k}=(k_x,k_y)$} that will be first used for wave-field extrapolation and then for aberration correction.}

\subsection{Propagator from the geophones to the focused basis }

The focusing operations described in {section \ref{redat} provide the FR matrix that plays a pivotal role in matrix imaging.} {We now show how this matrix can be obtained through simple matrix operations.}

Mathematically, the response between virtual sources and receivers is obtained from the response matrix at the surface through the Green's functions, that describe the propagation between each geophone and each point inside the medium using a wave velocity model. Switching between bases can be easily achieved by simple matrix products in the frequency domain. 
We first define the \review{plane-wave} propagator $\mathbf{T}_{{0}} (z,f)$, that enables a direct projection of the response matrix from the geophones' basis to the focused basis {$\bm{\rho}=(x,y)$} at each depth $z$. Each monochromatic response matrix \rev{$\overline{\mathbf{R}}_\mathbf{ss}(f)$} can be projected in the focused basis both at input and output by applying appropriate phase shifts associated with downgoing waves at input and upgoing waves at output to provide the FR matrix \rev{$\Rrrb$} (Fig.~\ref{chap3fig2}a). Under a matrix formalism, \rev{this operation can be written as follows:}
\begin{equation}
\label{Rrr1}
\Rrrb(z,f)=\mathbf{T}_{{0}}^{\dagger}(z,f)\times \rev{\overline{\mathbf{R}}_\mathbf{ss}}(f)\times\mathbf{T}_{{0}}^{*}(z,f).
\end{equation}
\revrita{where the symbols $*$ and $\dagger$ stands for phase conjugate and transpose conjugate, respectively.}

A model for the wave velocity distribution inside the medium is required. In this study, and since the horizontal EE cross-correlation functions are considered, only an \revrita{estimate} of the S-wave velocity is required. Unlike~\cite{blondel2018} and~\cite{touma2021} that considered a homogeneous P-wave velocity model, here a layered 1-D S-wave velocity model is used for the focusing process. A combination of two models derived by~\cite{kahraman2015}, for the first $5$ km, and by~\cite{karahan2001}, for deeper layers is displayed in Table~\ref{T1}. {Compared to an homogeneous model, such a} layered model will allow better time-depth conversion and will limit the aberration level in the subsurface image. \rev{Note, however, that our wave propagation model will not account for multiple reflections that could, in principle, take place between the interfaces of the different layers.}


\begin{table}[h!]
\centering
 \begin{tabular}{ccc}
\toprule
\multicolumn{1}{c}{ Layer {\#$i$}} &
\multicolumn{1}{c}{ Depth (km)} & \multicolumn{1}{c}{${{c}_{i}}\:(\textrm{m s}^{-1})$}\\
\midrule
  0 & 0 - 1 &  1700 \\
   1 & 1 - 3.5 &  2500 \\
   2 & 3.5 - 14 &  3200 \\
   3 & 14 - 26 &  3500 \\
   4 & 26 - 40 & 3600 \\
  5 & 40 - 60 & 4300 \\ [1ex] 
\bottomrule
\end{tabular}
\caption{1-D S-wave velocity model. {$c_{i}$ versus depth} following~\cite{karahan2001} and~\cite{kahraman2015}.}
\label{T1}
\end{table}

{In a layered medium,} the forward and backward extrapolation of the reflection matrix \rev{can be} performed through the decomposition of the wave-field into plane waves~\citep{berkhout1981}. Indeed, plane waves can be easily extrapolated by applying a simple phase shift. 
\rev{To that aim, w}e define the spatial transfer function, \rev{$F_i({\mathbf{k}},f)$}, that models the ballistic propagation of \rev{shear waves} through the $i^{th}$ layer\rev{:
\begin{equation}
F_i({\mathbf{k}},f)=
\left \{
\begin{array}{ll}
\exp \left (-i q_i \Delta z_i  \right)  & \, \mbox{for } \sqrt{k_x^2+k_y^2} <2\pi f/c_i  \\
0 & \,   \mbox{otherwise}
\end{array} 
\right.
,
\label{Gkz0}
\end{equation}}
with
\begin{equation*}
\rev{q_i}=\sqrt{\left (\frac{2\pi f}{c_i} \right ) ^2 -\rev{ k_x^2 - k_y^2}},
\end{equation*}
the \review{vertical} component of the wave vector \rev{$\mathbf{p}_i=(k_x,k_y,q_i)$} in the $i^{th}$ layer, $c_i$ the wave velocity in the $i^{th}$ layer of our model and $\Delta z_i$ its thickness. To propagate the plane waves from the surface to depth z, we define the wave-field extrapolator, $\mathbf{F}(z,f)=[F(\rev{\mathbf{k}},z,f)]$, as the product of the spatial transfer function\rev{s} of the $N$ layers above the considered depth as follows: 
 \begin{equation}
F({\mathbf{k}},z,f)=
\left \{
\begin{array}{ll}
\exp(-i \rev{q_N} (z-z_{N})) 
\rev{\prod_{i=1}^{N-1}F_i(\mathbf{k},f)} & \, \mbox{for } \rev{\sqrt{k_x^2+k_y^2}} <2\pi f/c_N \\
0 & \,   \mbox{otherwise.}
\end{array} 
\right.
,
\label{Gkz}
\end{equation}
where $z_i$ is the depth at which starts the $i^{th}$ layer. The \review{phase} propagator, $\mathbf{T}(z,f)=[T(\mathbf{s},\rev{\bm{\rho}},z,f)]$, can finally be expressed as follows:
 \begin{equation}
\mathbf{T}_{{0}}(z,f)=[\mathbf{P}_{0} \circ \mathbf{F}(z,f)] \times \mathbf{P}_{0}^{'\dag}.
\label{Gskz}
\end{equation}
\rev{where the symbol $\circ$ refers to the Hadamard product~\review{(\textit{i.e.} element wise matrix multiplication)}. In the \revrita{equation~\ref{Gskz}}, the term-by-term product arises because wave propagation in the plane wave basis is modeled by the scalar product of each plane wave component $P_0(\mathbf{k},\mathbf{s})$ with the overall spatial transfer function $F(\mathbf{k},z)$. The matrix multiplication stands for the inverse Fourier transform that enables \revrita{us} to project the propagated wave-field from the plane wave basis to the focused basis.}
 $\mathbf{{P}}_{0}^{'}=[{P}_{0}^{'}({\bm{\rho}},\rev{\mathbf{k}})]$ is actually the Fourier transform operator {linking the focused and plane wave bases.}
It connects the transverse wave vector $\rev{\mathbf{k}=( k_x,k_y)}$ of each plane wave to {the transverse coordinates $\bm{\rho}=(x,y)$ of each focusing point:}
\begin{equation}
\mathbf{P}_{0}^{'}\left({\bm{\rho} ,\mathbf{k}}\right) =\exp{\left(i \rev{\mathbf{k}} \cdot {\bm{\rho}}\right)}.
\label{T0'}
\end{equation}

To avoid aliasing during the change of basis between the plane wave and focused bases, a Shannon criterion should be respected. The \rev{transverse} {wave \rev{components, $k_x$ and $k_y$,} \review{are required} to fulfill the following condition:}
\begin{equation}
\rev{\sqrt{k_x^2+k_y^2}}<\rev{2\pi f/c_0}.
\label{Shannon1}
\end{equation}
The resolution \rev{$\delta k$} of the Fourier plane is conditioned by the size of the array $\mathcal{D} = 50$ km such that $\delta \rev{k}={2\pi /\mathcal{D}}$. 
{By properties of the Fourier transform, the transverse resolution $\delta {\rho}_0$ in the focal plane, that corresponds to the distance between the focusing points $\r$, is chosen to be $\sim \lambda /2 \sim 2 $ km to circumvent spatial aliasing in the focused basis.}

\rev{As shown by Eq.~\ref{Gkz}, wave components of spatial frequencies larger than the wavenumber $2\pi f/c_i$ cannot be transmitted through the i$^{th}$ layer. As a consequence, redatuming acts as a low-pass filter in the spatial frequency domain. To illustrate this phenomenon, one can back-project the focused reflection matrix in the geophone basis,
\begin{equation}
\label{K_prim}
\review{\overline{\mathbf{R}}'_{ss}(f)=\mathbf{T}_{0}(z,f) \times \Rrrb(z,f) \times \mathbf{T}_{0}^{\top}(z,f)},
\end{equation}
where the superscript $\top$ stands for matrix transpose. From $\overline{\mathbf{R}}'_{ss}(f)$, one can investigate the angular decomposition of the output wave-fields as previously done for the original response matrix $\overline{\mathbf{R}}_\mathbf{ss}$ in Figs.~\ref{fig10}(a) and (b) (Eq.~\ref{Rsk_0}). The result is displayed in Figs.~\ref{fig10}(c) and (d). The comparison with their original counterparts highlights the low pass-filter operated by redatuming: The surface wave component is discarded and only the shear waves associated with spatial frequencies $k<2\pi f/c_N$ are kept. Figure~\ref{fig10}(d) displays: (\textit{i}) a low-spatial frequency component associated with specular reflectors; (\textit{ii}) several off-axis bright spots associated with peculiar single scattering events at depth $z$ and; (\textit{iii}) a diffuse background that is difficult to interpret at this stage since it can be due to random single scattering, multiple scattering or noise resulting from the imperfect convergence of cross-correlations towards the Green's function. To enhance the single scattering contribution with respect to the other undesirable components for imaging, the idea is now to perform a time gating operation to enhance the single scattering contribution.}

\subsection{Broadband focused reflection matrix}

\rev{Equation \ref{Rrr1} simulates focused beamforming for both downgoing (input) and upgoing (output) shear waves at each frequency. Each spectral component of the wave-field can then be recombined to provide a broadband focused reflection matrix $\Rrrb(z)$:}
\begin{equation}
\label{Rrr}
{\Rrrb}(z)=\int^{f_{2}}_{f_{1}} df \rev{\Rrrb(z,f)}, 
\end{equation}
with $f_1=0.1$ Hz and $f_2=0.5$ Hz. This operation amounts to \revrita{performing} an inverse Fourier transform at time $t = 0$ over the frequency band [0.1 0.5] Hz. \rev{Note that this inverse Fourier transform is only performed over positive frequencies in order to have access to both the amplitude and phase of the wave-field.} {\rev{The time origin here corresponds to the ballistic time in the focused basis. Equation~\ref{Rrr} thus corresponds to a time gating operation} that tends to select singly-scattered waves associated with a scattering event in the focal plane. } 
{Each coefficient ${{R}}({\bm{\rho}_\textrm{out}},{\bm{\rho}_\textrm{in}},z)$ of ${\Rrr}(z)$} contains the \rev{complex} wave-field that would be recorded by a virtual geophone located at $\rout=({\bm{\rho}_\textrm{out}},z)$ if a virtual source at $\rin=({\bm{\rho}_\textrm{in}},z)$ emits a pulse of length $\delta t= \Delta f^{-1}$ at the central frequency $f_0$, with $\Delta f=f_2-f_1 = 0.4$ Hz.


\begin{figure*}
 \centering
    \includegraphics[width=12cm]{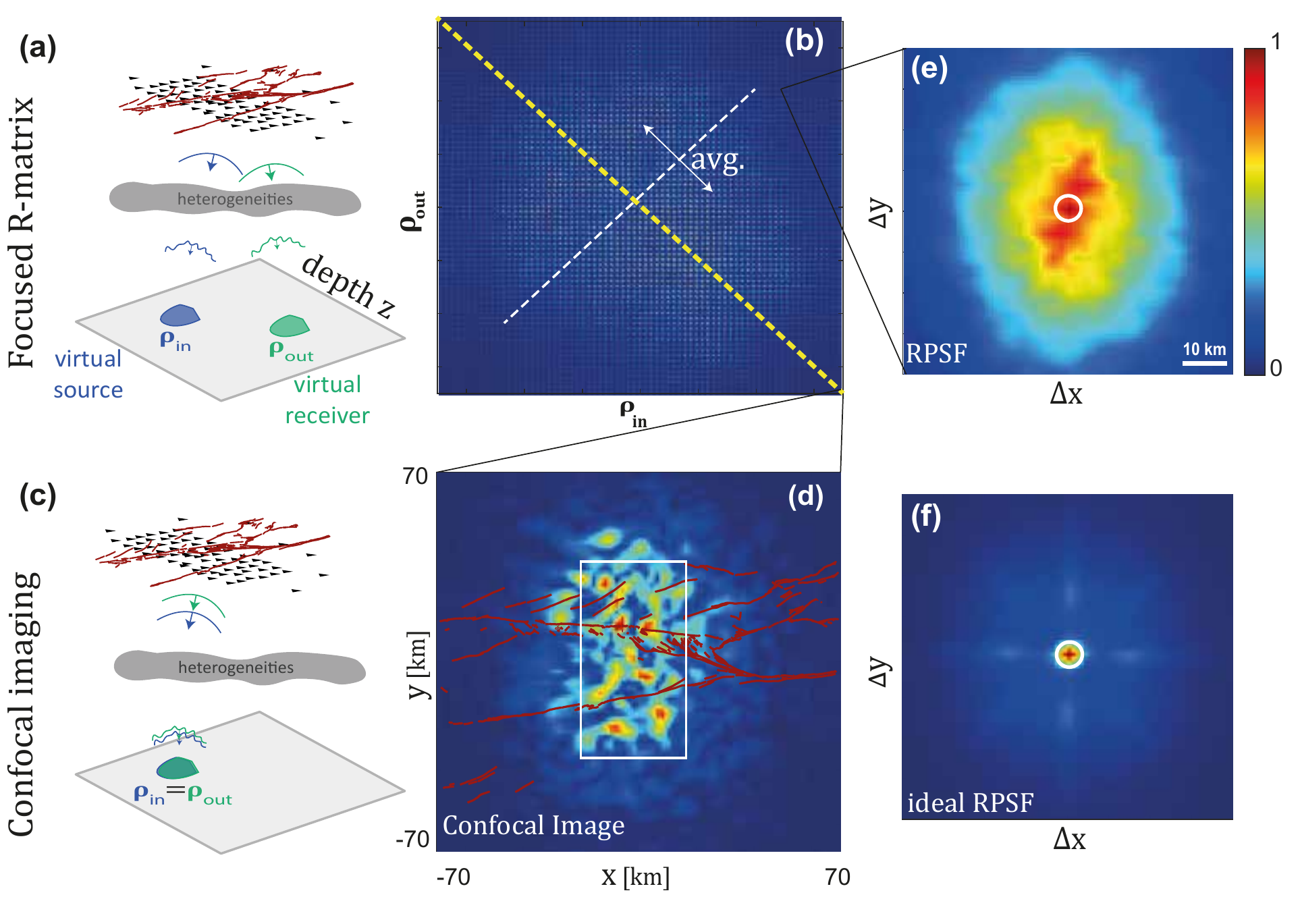}
 \caption{{Focused reflection matrix. (a) The response matrix $\mathbf{K}$ is projected onto a focused basis at each depth $z$ (Eq.~\ref{Rrr}), thereby synthesizing a set of virtual sources ($\bm{\rho}_\textrm{in}$) and receivers ($\bm{\rho}_\textrm{out}$) scanning laterally the field-of-view. In presence of fluctuations in the seismic velocity, focused waves are distorted while travelling from the surface to the plane, thereby enlarging and distorting the virtual geophones. (b) This effect gives rise to a off-diagonal spreading of backscattered energy in the focused reflection matrix {$\Rrr(z)$} shown here at depth $z =25$ km. (c) The corresponding intensity profile, averaged over whole the field-of-view, provides the so-called RPSF $I$ (Eq.~\ref{Ilocal}). The white circle represents the diffraction-limited  transverse resolution ($\delta \rho_0 \sim 6$ {km}) at the considered depth. (d) The ideal RPSF that would be obtained in absence of aberrations is shown for comparison (e) Confocal image $\mathcal{I}$ (Eq.~\ref{image}) built from the diagonal of {$\Rrr$}. { The white box represents the dimensions of the rectangular array of geophones and the red lines represent the NAFZ fault traces at the surface. {(f) The confocal image corresponds to a simultaneous focusing process at input and output ($\mathbf{r}_\textrm{in}=\mathbf{r}_\textrm{out}$).}}
 In panels (b)-(e), the color scale refers to the scattering intensity. It is normalized by the maximum value of the scattering energy at the considered depth.} }
\label{chap3fig2}
  \end{figure*}
  
The FR matrix can be expressed theoretically as follows~\citep{lambert2020,lambert2021refle,touma2021},
\begin{equation}
\label{RrrMatrix}
\rev{\Rrr}(z)={\mathbf{H}^{\top}}(z)\times \mathbf{\Gamma}(z) \times {\mathbf{H}}(z),
\end{equation}
\review{
\rev{which yields, in terms of matrix coefficients,}}
\begin{equation}
\label{Rrr_coef2}
{{R}}({\bm{\rho}}_\textrm{out},{\bm{\rho}}_\textrm{in},z)=\int \review{d\bm{\rho} } {H}({\bm{\rho}},{\bm{\rho}}_\textrm{out},z ) \gamma({\bm{\rho},z}) {H}({\bm{\rho}},{\bm{\rho}}_\textrm{in},z).
\end{equation}
\review{The matrix $\mathbf{\Gamma}$ describes the scattering process in the focused basis. In the single scattering regime, this matrix is diagonal and its coefficients correspond to the \revrita{subsurface} reflectivity $\gamma({\bm{\rho}},z)$ at depth $z$. } \rev{$\mathbf{H}(z)$} is the focusing matrix whose coefficients ${H(\bm{\rho},\bm{\rho}_{\textrm{in}/\textrm{out}},z)}$ correspond to the point spread functions \review{(PSFs)} of the redatuming process. \review{These PSFs represent the spatial amplitude distribution of the focal spots} \rev{for each focusing point} $\r_{\textrm{in}/\textrm{out}}{=(\bm{\rho}_{\textrm{in}/\textrm{out}},z)}$. 
{They thus} account for the lateral extent of each vitual source/detector at $\r_{\textrm{in}/\textrm{out}}$.

An example of the broadband FR matrix {$\Rrr$} is shown at depth $z = 25$ km in Fig.~\ref{chap3fig2}{b}.
$\rev{\Rrr}$ is a four-dimension matrix concatenated in 2D as a set of blocks~\citep{blondel2018}. \rev{If the wave velocity model was correct, the PSFs of the redatuming process would be close to be point-like [$H(\bm{\rho},\bm{\rho}_{\textrm{in}/\textrm{out}},z) \simeq \delta (\rho-\bm{\rho}_{\textrm{in}/\textrm{out}})$, with $\delta$ the Dirac distribution] and the focused reflection matrix ${\Rrr}$ almost diagonal [${R}({\bm{\rho}}_\textrm{out},{\bm{\rho}}_\textrm{in},z)\simeq \gamma ({\bm{\rho}}_\textrm{in}) \delta (\bm{\rho}_{\textrm{in}}-\bm{\rho}_{\textrm{out}}) $, see Eq.~\ref{Rrr_coef2}].} Here, {the backscattered energy is far from being concentrated along the diagonal of {$\Rrr$}, \rev{which is a manifestation of the gap between our layered wave velocity model (Table~\ref{T1}) and the real shear wave velocity distribution in the \revrita{subsurface}.} } 

%

\subsection{\rev{Confocal image}}

{Nevertheless, one can try to build an image of the medium reflectivity at each effective depth $z$ by considering the diagonal elements of the FR matrix, \review{i.e where the virtual sources and receivers coincide} ($\bm{\rho}_\textrm{in}=\bm{\rho}_\textrm{out}=\bm{\rho}_c$, see Fig.~\ref{chap3fig2}{c}).} It yields the \rev{so-called} \textit{confocal} image:
\begin{equation}
\label{image}
\mathcal{I}\left(\bm{\rho}_c,z\right) = {{R}}\left(\bm{\rho}_c,\bm{\rho}_c,z\right).
\end{equation}
Fig.~\ref{chap3fig2}{e} shows the resulting 2D image $\mathcal{I}$ at $z = 25$ km retrieved from the diagonal of the FR matrix in Fig.~\ref{chap3fig2}{b}. \rev{By injecting Eq.~\ref{Rrr_coef2} into the last equation, $\mathcal{I}$ can be expressed as the transverse convolution between the medium reflectivity and the confocal PSF $H^2$, such that:
\begin{equation}
\label{image2}
\mathcal{I}\left(\bm{\rho}_c,z\right) = \int d\bm{\rho}   \gamma({\bm{\rho},z}) {H}^2({\bm{\rho}},{\bm{\rho}}_c,z).
\end{equation}
Such an image is thus a reliable estimator of the reflectivity at depth $z$ only if the wave velocity model is close to reality. In this ideal case, the spatial extent of the PSF is only limited by diffraction and the transverse resolution is given by}~\citep{born}: 
\begin{equation}
\label{deltax}
{\delta {\rho_0}=\lambda / (2 \sin \theta)}
\end{equation}
where $\theta=\arctan (\mathcal{D} /2z)$ is determined by the size of the array $\mathcal{D} = 50$ km, and corresponds to the maximum angle under which a focusing point sees the geophones' array. 

\begin{figure*}
 \centering
 \includegraphics[width=\textwidth]{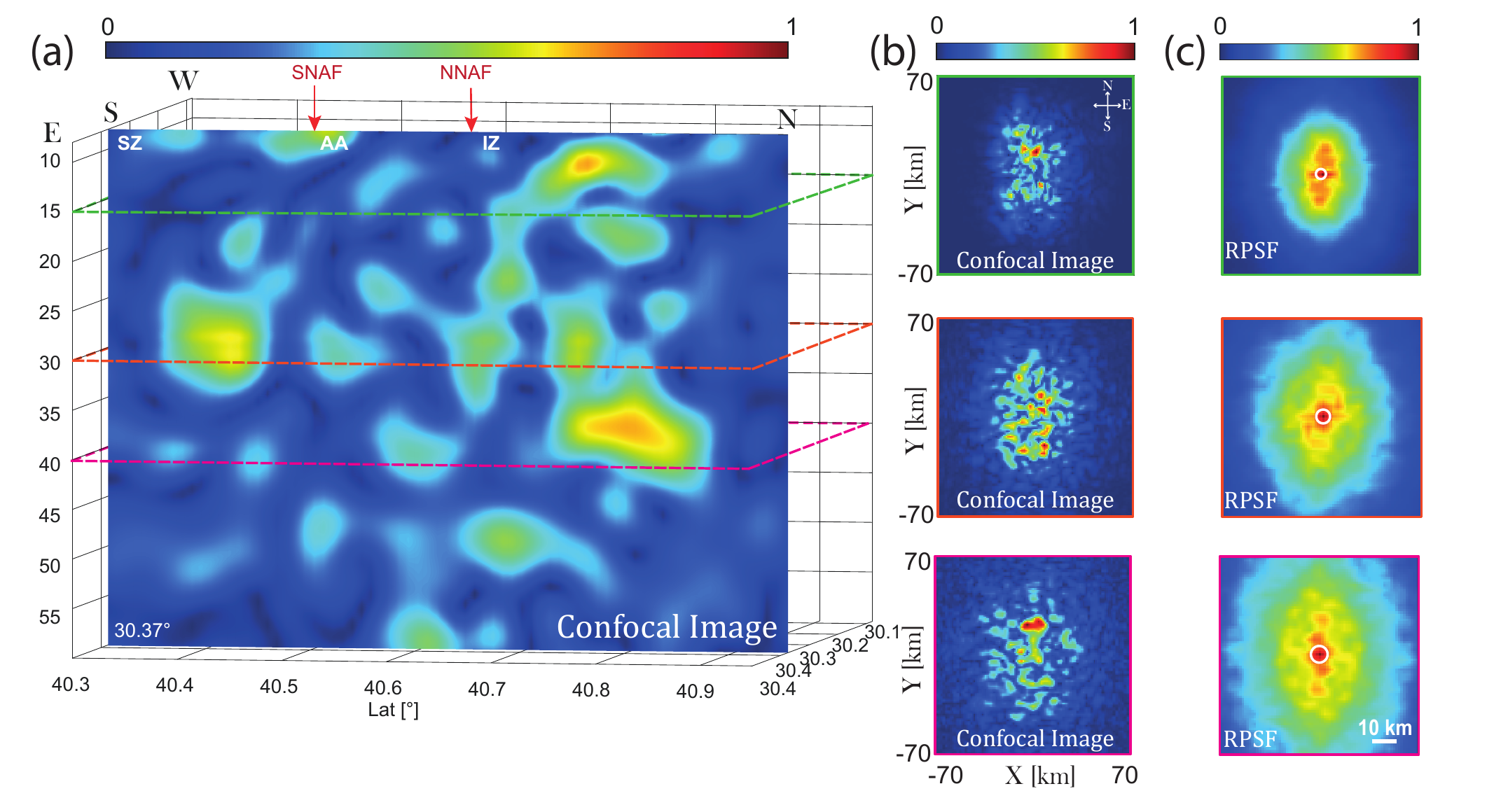}
 \caption{\rev{Original confocal image of NAFZ.} (a) Vertical North-South cross-section at $30.37\degree$E. The North-South profile is oriented perpendicular to the fault traces. The location of the profile is shown in Fig.~\ref{chap3fig1}a. The locations of the southern (SNAF) and northern (NNAF), and the major crustal blocks (SZ: Sakarya zone, AA: Armutlu-Almacik and IZ: Istanbul zone) are labeled. The color scale refers to the scattering intensity. It is normalized by the maximum value of the scattering energy inside the volume. (b) Depth slices retrieved from the 3D scattering volume at $z=15, 30$ and $40$ km with (c) their corresponding {RPSFs}.}
\label{chap3fig5}
  \end{figure*}

By stacking the confocal image computed at each depth $z$, a 3D image of the reflectivity can be obtained. The cross-section at Lon $30.37\degree $ is displayed in {Fig.~\ref{chap3fig5}\rev{a}. 2D \textit{confocal} images at $z = 15, 30$ and $40$ km are also shown in Fig.~\ref{chap3fig5}\rev{b}}. Unlike the transverse resolution, the axial resolution $\delta z$ is limited by the frequency bandwidth: $\delta z \sim c /\Delta f\sim 8.7$ km, with $c$ the shear wave velocity at the considered depth. The sections of the 3D image displayed in {Figs.~\ref{chap3fig2}e,~\ref{chap3fig5}a and~\ref{chap3fig5}b} show a greater reflectivity in the central part of the field-of-view, i.e right below the geophones' array, but no direct correlation can be found between the image and the location of the fault strands. \rev{In fact, as we will see now, lateral wave speed heterogeneities strongly degrade the transverse resolution of the redatuming process and induce strong aberrations in the confocal image. \revrita{This image} is thus not a reliable estimator of the medium reflectivity \rev{at this stage} and cannot be interpreted.}

{\subsection{Quantification of aberrations}}

The FR matrix {can provide more than a confocal image since its off-diagonal elements can lead to a quantification of aberrations.}
{{To that aim, a} relevant observable is the distribution of the backscattered intensity around a common midpoint point \rev{$(\bm{\rho}_m,z)$} as a function of the relative position\rev{, $\Delta {\bm{\rho}}=\bm{\rho}_\textrm{out}-\bm{\rho}_\textrm{in}$,} between the input and output focusing points~\citep{lambert2020,touma2021}:}
\begin{eqnarray}
  I(\Delta {\bm{\rho}},\rev{\bm{\rho}}_m,z ) & = & \rev{ |\rev{{R}}(\rev{\bm{\rho}_m} - \Delta {\bm{\rho}} /2,\rev{\bm{\rho}_m} + \Delta {\bm{\rho}}/2{,z})|^2} .
  \label{Ilocal}
\end{eqnarray}
\rev{In the following, we will refer to this quantity as the reflection point spread function (RPSF).}

\rev{To express this quantity theoretically, we first make a \rev{local isoplanatic} approximation in the vicinity of each point $(\bm{\rho}_m,z)$~\citep{lambert2021refle}. Isoplanicity means here that waves which focus in this region are assumed to have travelled through approximately the same areas of the medium, thereby undergoing identical phase distortions. The PSF can then be considered to be spatially invariant within this local region.
Mathematically, 
{this} means that, in the vicinity of each point $(\bm{\rho}_m,z)$, the spatial distribution of the PSF, $H(\bm{\rho},\bm{\rho}_\text{in/out},z)$, only depends {on} the relative distance between the point $\bm{\rho}$ and the focusing point $\bm{\rho}_{\text{in/out}}$. This leads us to define a local spatially-invariant PSF $H_L$ around each common mid-point $(\bm{\rho}_m,z)$ such that:
\begin{equation}
\label{isoplanatic}
H(\bm{\rho},\bm{\rho}_\text{in/out},z) =  H_L(\bm{\rho}-\bm{\rho}_\textrm{in/out},\bm{\rho}_m,z).
\end{equation}}

Under this local isoplanatic assumption, the RPSF can be derived analytically in different scattering regimes. On the one hand, for large reflectors such as horizontal interfaces between geological units, the medium reflectivity can be assumed as locally constant and the RPSF is given by \revrita{(see Supplementary Section S1)}:
\begin{equation}
I (\Delta {\bm{\rho}},\bm{\rho}_m,z )  =  |\gamma(\bm{\rho}_m,z)|^2 \left |H_L \stackrel{\Delta {\bm{\rho}}}{\circledast} H_L \right |^2 (\Delta {\bm{\rho}},\bm{\rho}_m,z) .
\end{equation}
\review{where the symbol ${\circledast}$ denotes convolution.} On the other hand, for diffuse scattering, the medium reflectivity can be considered, in first approximation, as randomly distributed. Under that assumption, the mean RPSF is then proportional to the convolution between the incoherent output and input local PSFs, independently from the medium's reflectivity~\citep{lambert2020} (see Supplementary Section S1):
\begin{equation}
\langle I(\Delta {\bm{\rho}},\bm{\rho}_m,z)\rangle \propto \left[ \left| H_L \right|^{\rev{2}}\stackrel{\Delta {\bm{\rho}}}{\circledast} \left|H_L \right|^{\rev{2}}\right] \left(\Delta {\bm{\rho}},\bm{\rho}_m,z\right)
\end{equation}
\rev{where the symbol $\langle \cdots \rangle$ stands for an ensemble average. Whatever the scattering regime, the spatial extension of the RPSF is thus roughly equal to the lateral dimension of the PSF. If we assume a Gaussian PSF, this equality is \revrita{strict}. The RPSF is thus a direct indicator of the focusing quality} and its spatial extent directly provides an estimation of the local transverse resolution of the confocal image.

Fig.~\ref{chap3fig2}{c} displays the {RPSF} averaged over the whole  field-of-view at depth $z=25$ km.
{For sake of comparison, Fig.~\ref{chap3fig2}{d} shows the ideal (i.e diffraction-limited) {RPSF} that would be obtained in absence of aberrations.}
{The comparison between Figs.~\ref{chap3fig2}{c and d} highlights the impact of aberrations \review{resulting from} the mismatch between the wave velocity model of Table~\ref{T1} and the real wave speed distribution. Indeed, the full width at half maximum $w$ of the intensity profile is increased by a factor $\sim 6$ compared to its diffraction-limited value (white circle in Fig.~\ref{chap3fig2}{d, Eq.~\ref{deltax}}) at depth  $z=25$ km. \revrita{This effect} explains the blurred aspect of the confocal image displayed in Fig.~\ref{chap3fig2}{e} at the same depth.}
{The {impact} of aberrations is also \revrita{illustrated in} {by Fig.~\ref{chap3fig5}\rev{c} that displays the depth evolution of the RPSF} inside the Earth. As \revrita{with} the diffraction-limited resolution (Eq.~\ref{deltax}), the transverse extension of the RPSF also increases with $z$ but it shows a much larger extension. } 


In the following we will show how matrix imaging can restore an optimal resolution for this image.
\vspace{5 mm}

\section{{Exploiting the input-output angular correlations of the wave-field: The CLASS algorithm}}
\label{sec:CLASS}

{In order to compensate for aberrations, the reflection matrix can be first projected in the plane wave basis:} 
\begin{equation}
\label{Rkk}
\mathbf{R_{kk}}(z)=\mathbf{P}_{0}^{'} \times {\Rrr}(z) \times \mathbf{P}_{0}^{'\top},
\end{equation}
\rev{Using Eq.~\ref{T0'}, the last equation can be rewritten, in terms of matrix coefficients, as a double spatial Fourier transform:
\begin{equation}
R(\mathbf{k}_\textrm{out},\mathbf{k}_\textrm{in},z)=\sum_{\bm{\rho}_\textrm{out}}\sum_{\bm{\rho}_\textrm{in}} e^{-i \mathbf{k}_\textrm{out} . \bm{\rho}_\textrm{out}} R(\bm{\rho}_\textrm{out},\bm{\rho}_\textrm{in},z) e^{-i \mathbf{k}_\textrm{in} . \bm{\rho}_\textrm{in}}
\end{equation}}
Each coefficient of the matrix $\mathbf{R_{kk}}(z)=[R(\mathbf{k}_\textrm{out},\mathbf{k}_\textrm{in},z)]$ \rev{thus} contains the 
medium response between input and output transverse wave vectors $\k_\textrm{in}$ and $\k_\textrm{out}$. \rev{By injecting Eq.~\ref{RrrMatrix} into Eq.~\ref{Rkk},} the matrix $\Rkk$ can be expressed as follows:
\begin{equation}
\label{RkkMatrix}
\mathbf{R_{kk}}(z)={\mathbf{T}}(z) \times \mathbf{\Gamma}(z) \times {\mathbf{T}}^{\top}(z),
\end{equation}
\rev{where $\mathbf{T}(z)= \mathbf{P}^{'}_{0} \times \mathbf{H}(z)$
is the transmission matrix describes plane wave propagation between the focused and the plane wave bases. Its coefficients $T(\k,\bm{\rho},z)$ correspond to the angular decomposition of the wave-field produced at the Earth surface for a point-like virtual source located at $(\bm{\rho},z)$. This matrix is \revrita{critical} for imaging since its inversion can provide a direct access to the \revrita{subsurface} reflectivity, without relying on a precise wave velocity model.}

\rev{As a first step towards the estimation of $\mathbf{T}$, we can go one step further in the isoplanatic approximation (Eq.~\ref{isoplanatic}) by assuming a full transverse-invariance of the PSF across the field-of-view. This leads us to define a laterally-invariant PSF $H_I$, such that: 
\begin{equation}
\label{iso}
H_L(\bm{\rho}-\bm{\rho}_\text{in/out},\bm{\rho}_m,z) \simeq H_I(\bm{\rho}-\bm{\rho}_\text{in/out},z).
\end{equation}
\rev{This strong assumption means that wave speed heterogeneities are} modelled by a phase screen of transmittance $ \mathbf{\Tilde{H}}_I =[\tilde{H}_I(\k,\rev{,z})]$ in the plane wave basis, such that } ${\mathbf{T}}\rev{(z)}= \mathbf{\Tilde{H}}_{\rev{I}}\rev{(z)}\circ \mathbf{P}^{'}_{0}$, where  $\tilde{H}_I(\k\rev{,z}) =  \int \review{d\bm{\rho} } H_{\rev{I}}({\bm{\rho}},z) e^{-i  {\mathbf{k}.\bm{\rho}}}$ is the Fourier transform of the spatially-invariant PSF ${H}_{\rev{I}}  ({\bm{\rho}}\rev{,z}) $. \rev{The aberration transmittance $\mathbf{\Tilde{H}}_I$ grasps the phase distortions undergone by downgoing and upgoing wave-fields during their travel between the Earth surface and the focal plane at effective depth $z$. }

\begin{figure*}
\centering
\includegraphics[width=\textwidth]{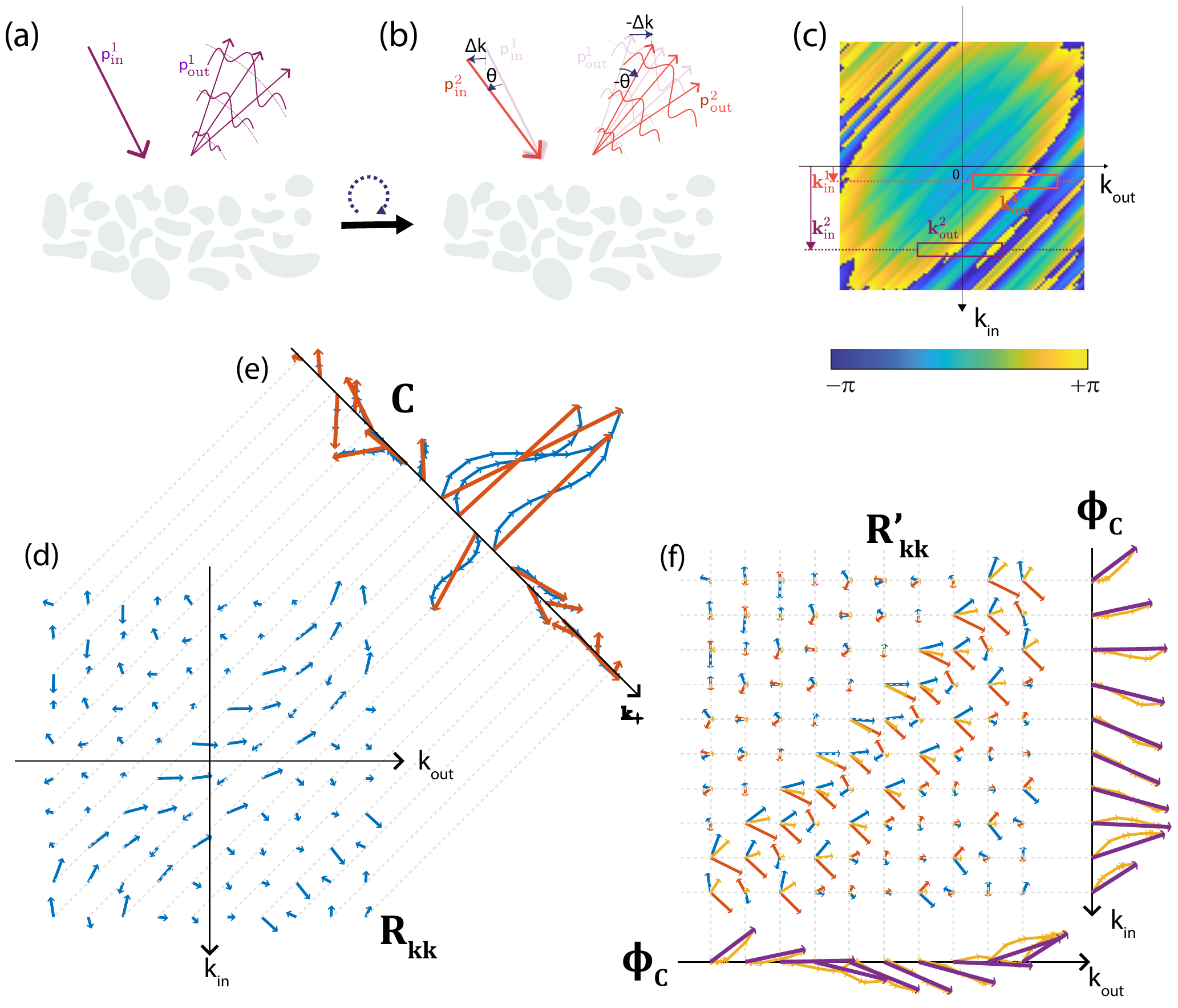}
\caption{\rev{Principles of the memory effect and CLASS algorithm. (a) When
an incident plane wave of wave vector $ \mathbf{p}_\textrm{in} $ insonifies a scattering medium, the reflected wave-field exhibits a random speckle pattern. (b) When this incident wave-field is rotated by an angle $\theta$, the reflected wave-field is shifted by the opposite angle $-\theta$: This is the so-called memory effect. (c) This phenomenon results in a deterministic coherence along the antidiagonals  ($\mathbf{k}_\textrm{in}+\mathbf{k}_\textrm{out}=$constant) of the reflection matrix $\mathbf{R_{kk}}$ expressed in the plane wave basis. The phase of a matrix $\mathbf{R}_{kk}$ is displayed for sake of illustration. This matrix has been obtained from an ultrasound experiment performed on a medium of random reflectivity (acoustic phantom) in the conditions described by~\cite{lambert2020}. (d) Same matrix as in (c) but in presence of aberrations. Each complex coefficient of $\Rkk$ is here represented by blue arrows using a Fresnel diagram. (e) The first step of CLASS (Eq.~\ref{C}) consists in a coherent sum of coefficients lying along the same antidiagonal to estimate the spatial frequency spectrum $\tilde{\gamma}$ of the medium reflectivity. Each coefficient of the resulting vector $\mathbf{C}$ is represented by a red arrow in the complex plane. (f) The second step of CLASS consists in a compensation of $\Rkk$ by $\mathbf{C}^*$ to compensate for the phase of reflectivity spectrum $\tilde{\gamma}$ (Eq.~\ref{Rtmp}). The coefficients of the resulting matrix $\Rkkp$ are depicted with orange arrows. A sum over lines or columns of $\Rkkp$ provides the isoplanatic aberrated wave-front $\exp (i \bm{\phi}_c)$ represented by purple arrows (Eq.~\ref{phiout}).}}
\label{MemoryEffect}
\end{figure*}

\rev{Under this full isoplanatic approximation (Eq.~\ref{iso}),} a theoretical expression of $\Rkk$ can be derived in the single scattering regime~\citep{Lambert2020b}:
\begin{equation}
\label{Rkkcoeff}
R\left(\kout , \kin , z\right)=\tilde{H}_{\rev{I}}\left(\kin \rev{, z} \right) \tilde{\gamma}\review{\left(\kin +\kout , z\right)} \tilde{H}_{\rev{I}}\left(\kout \rev{, z} \right),
\end{equation}
where $\tilde{\gamma}\left(\k, z\right)=\int d {\bm{\rho}}\gamma({\bm{\rho}}, z) \exp \left(-i \k \cdot {\bm{\rho}}\right)$ is the 2D Fourier transform of the medium's reflectivity $\gamma({{\bm{\rho}},z})$ \rev{at each depth $z$}.
In absence of aberrations $\left (\tilde{H} (\mathbf{k}\rev{,z}) \equiv 1\right )$, the reflection matrix {expressed in the plane wave basis} exhibits a deterministic coherence along its antidiagonals ($\kin+\kout$=constant, see Fig.~\ref{MemoryEffect}c): \rev{$R(\kin,\kout,z)=\tilde{\gamma}(\kin+\kout,z)$}~\citep{aubry2,Kang2015}. This peculiar property is a manifestation of a phenomenon called the memory effect in wave physics~\citep{freund1988,shahjahan}. \rev{When an incident plane wave ($\mathbf{k}_\textrm{in}$) illuminates a scattering medium, it gives rise to a reflected wave-field ($\mathbf{k}_\textrm{out}$) that exhibits a speckle feature (Fig.~\ref{MemoryEffect}a) due to the random interference between partial waves induced by each scatterer lying at depth $z$. When this incident plane wave is rotated by a certain angle ($\mathbf{k}_\textrm{in}+\Delta \mathbf{k}$), the reflected wave-field is tilted in the opposite direction ($\mathbf{k}_\textrm{in}-\Delta \mathbf{k}$, see Fig.~\ref{MemoryEffect}b). This correlation between the downgoing and upgoing wave-fields accounts for the deterministic coherence along the antidiagonals of $\Rkk$ (Fig.~\ref{MemoryEffect}c).} \rev{However, in the present case, this property is not checked because of the phase distortions undergone by downgoing and upgoing wave-fields induced by wave speed heterogeneities. Mathematically, this is accounted by the phase screen $\tilde{H}_{\rev{I}}\left(\mathbf{k},z \right) $ in Eq.~\ref{Rkkcoeff} that breaks the correlation between coefficients lying along the same antidiagonal (Fig.~\ref{MemoryEffect}d). }

The principle of the CLASS algorithm~\citep{kang2017, choi2018, yoon2019} consists in restoring this coherence by applying a phase correction {, $\exp\left [ - i\phi_C(\mathbf{k}\rev{,z}) \right ]$}, {to the input and output of $\Rkk$:
{\begin{equation}
\label{RkkcCLASS}
R^{(C)}\left(\kout, \kin \rev{, z} \right) =  e ^{-i \phi_C \left(\kout \rev{,z} \right)}  R\left(\kout, \kin \rev{,z} \right)  e ^{-i \phi_C \left(\kin \rev{,z} \right)},
\end{equation}}
\rev{where $\phi_C$ is the estimator of the aberration phase law whose phase conjugate maximizes the coherence along the antidiagonals of $\Rkk$. To compute $\phi_C$, the first step is to perform a coherent sum of $\Rkk$'s coefficients along each of its antidiagonals [see Fig.~\ref{MemoryEffect}e]:
\begin{equation}
\label{C}
{C} \left(\kpl \rev{,z}  \right)=\sum_{\kout } R\left(\kout, \kpl - \kout \rev{, z} \right)
\end{equation}
\rev{with $\kpl=\kin+\kout$. As shown in Supplementary Section S2, ${C}(\kpl,z)$ is a rough estimator for the spatial frequency spectrum $\tilde{\gamma}(\kpl,z)$ of the medium reflectivity (see Supplementary Section S2). The second step consists in performing the Hadamard product~\review{(element-wise product)} between the phase conjugate of the vector $\mathbf{C}$ and the matrix $\Rkk$ (Fig.~\ref{MemoryEffect}f):}
\begin{equation}
\label{Rtmp}
{R'}\left(\kout, \kin, z \right)= R\left(\kout, \kin  \rev{, z}  \right) C^{*}\left(\kout + \kin  \rev{, z} \right).
\end{equation}
\rev{The last operation amounts to compensate for the phase fluctuations of the reflectivity spatial frequency spectrum $\tilde{\bm{\gamma}}$ in $\Rkk$. The isoplanatic phase distortion $\tilde{\mathbf{H}}_I$ is finally estimated by summing the columns of the compensated matrix $\mathbf{R}'_{\mathbf{kk}}$ (see Fig.~\ref{MemoryEffect}f)}
\begin{equation}
\label{phiout}
\phi_{C}\left(\kout  \rev{, z}  \right)=\arg \left[\sum_\kin R'\left(\kout, \kin  \rev{, z}  \right)\right],
\end{equation}}
\rev{The phase conjugate of the resulting wave-front, $\exp \left [-i \phi_C(\mathbf{k}\rev{,z} )\right ]$, tends to realign in phase the coefficients lying on the same antidiagonal of $\Rkk$ (Eq.~\ref{RkkcCLASS}). As shown in Supplementary Section S2, this phase realignment in the plane wave basis is equivalent to a maximization of the confocal intensity in the focused basis. }  }

\begin{figure*}
 \centering
 \includegraphics[width=\textwidth]{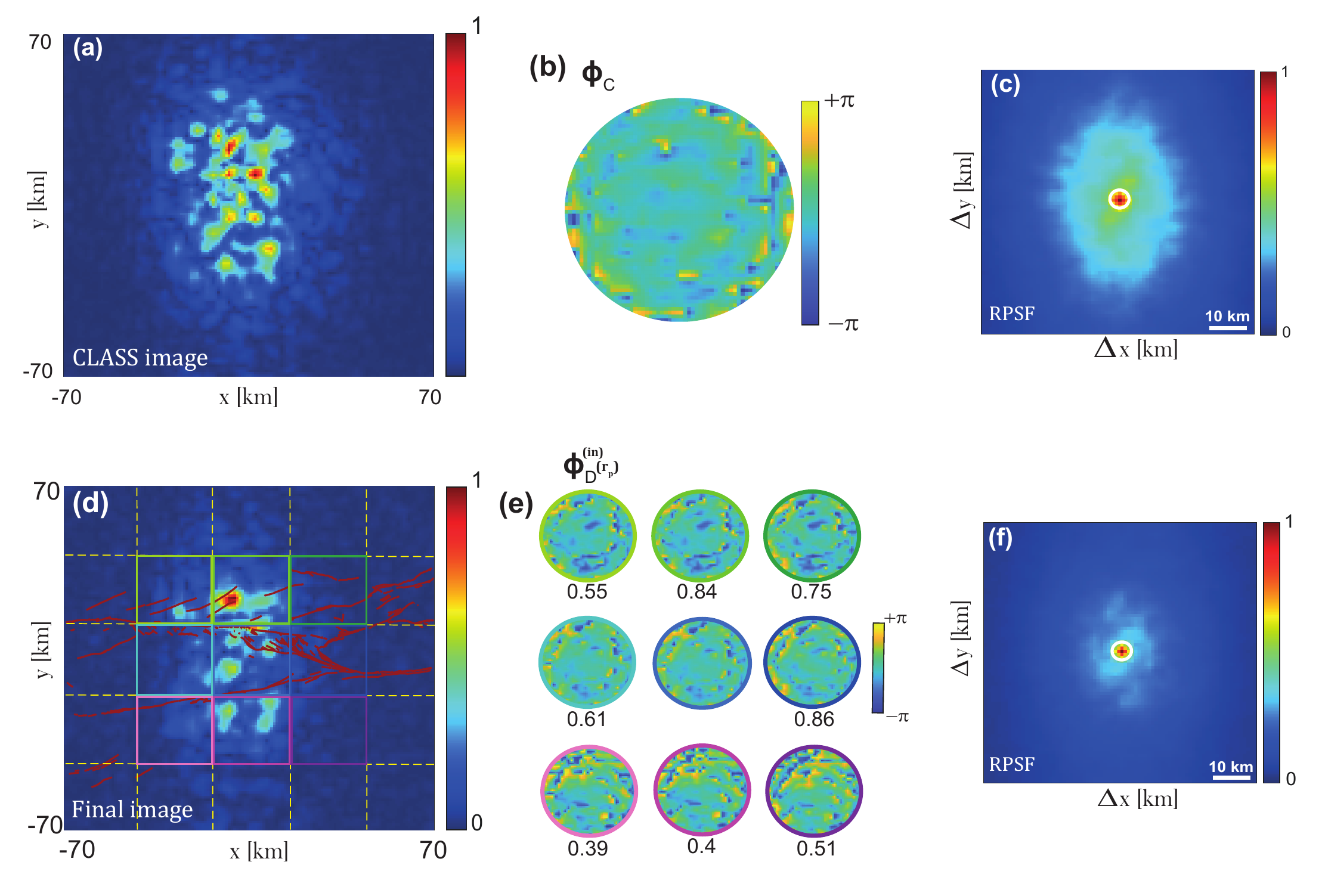}
 \caption{{Aberration correction process {at $z=25$ km}.
 (a) Confocal image obtained after applying {the conjugate of (b) the CLASS phase law $\phi_C$ computed at this depth. \review{(c)} RPSF obtained after CLASS correction}.}
 (d) Confocal image obtained after performing four iteration steps of the {distortion matrix process}. The red lines represent the NAFZ fault traces at the surface. The yellow dashed lines delineate the regions {over which a local aberration phase law $\phi(\mathbf{k},\mathbf{r})$ has been estimated. (e) Corresponding input aberration phase laws $\phi_\textrm{in}(\mathbf{k},\mathbf{r})$ obtained at the end of the process. The correlation coefficients between the corresponding aberration transmittances and the central one are displayed below each phase mask.
 (f) RPSF at the end of the matrix imaging process.} 
}
\label{chap3fig4}
  \end{figure*}

\rev{The corresponding FR matrix $\Rrr^{(C)}$} can be deduced as follows:
\begin{equation}
\label{RrrcCLASS}
\Rrr^{(C)}(z)=\mathbf{P}_{\mathbf{0}}^{\prime\dagger}(z) \times \Rkk^{(C)}(z) \times \mathbf{P}_{\mathbf{0}}^{^{\prime*}}(z)
\end{equation}
A corrected confocal image is extracted from the diagonal of $\Rrr^{(C)}(z)$ and displayed in Fig.~\ref{chap3fig4}a at depth $z=25$ km . 
{It should be compared with the original image shown in Fig.~\ref{chap3fig2}{e}. While the latter one displays a random-like feature, the corrected image reveals a greater reflectivity in the North that can be correlated with the expected damage around the Northern branch of the fault. The comparison between these two images illustrates the benefit of the correction process.} 
The gain in resolution can be assessed \rev{by averaging the RPSF }(Eq.~\ref{Ilocal}) over the whole field-of-view (see Fig.~\ref{chap3fig4}c). Compared to the original {RPSF} displayed in Fig.~\ref{chap3fig2}{c}, {we can notice that a large component of the off-diagonal energy has been brought back to the confocal lobe (white circle)}. {The resolution $ w$ is reduced from $40$ km to $8$ km but it is still larger than the diffraction-limited resolution {($\delta \rho_0 \sim 6$ km at the considered depth)}.} A diffuse component subsists and can be explained by the spatially-varying residual aberrations, {$\delta \tilde{H}_{\rev{L}}(\mathbf{k},\bm{\rho}_m,z)$}, that have not been compensated by the CLASS algorithm, such that {$\delta \rev{\tilde{H}_L}(\mathbf{k}\rev{,\bm{\rho}_c,z})=\rev{\tilde{H}_L}(\mathbf{k}\rev{,\bm{\rho}_c,z})e^{-i \phi_C (\mathbf{k},z)}$}.


{A local compensation of higher order aberrations is thus required.}
This issue is handled in the following section by {investigating the reflection matrix and its distorted component between the focused and plane wave bases. }

\section{Matrix approach for adaptive focusing: {The local distortion matrix}}

The distortion matrix $\mathbf{D}$ was already introduced in ultrasound~\citep{Lambert2020b,lambert2021distor}, optics~\citep{Badon2019,Najar2023} and seismology~\citep{touma2021}. Several applications proved the efficiency of this matrix in overcoming aberrations and improving the image quality.
Recent works in seismology~\citep{touma2021} and optics~\citep{Badon2019} have shown that for certain scattering regimes (specular reflectors or sparse scattering), there was a one-to-one association between the eigenstates of $\mathbf{D}$ and the isoplanatic patches present in the field-of-view. 
{Here, this property does not hold because the NAFZ subsurface exhibits a continuous \rev{but fluctuating} reflectivity (see Supplementary Section S3). In this \rev{scattering} regime,} local distortion matrices should be considered over restricted areas in which the isoplanatic hypothesis is ideally fulfilled~\citep{lambert2021distor,Najar2023}.

\begin{figure*}
 \centering
 \includegraphics[width=0.8\textwidth]{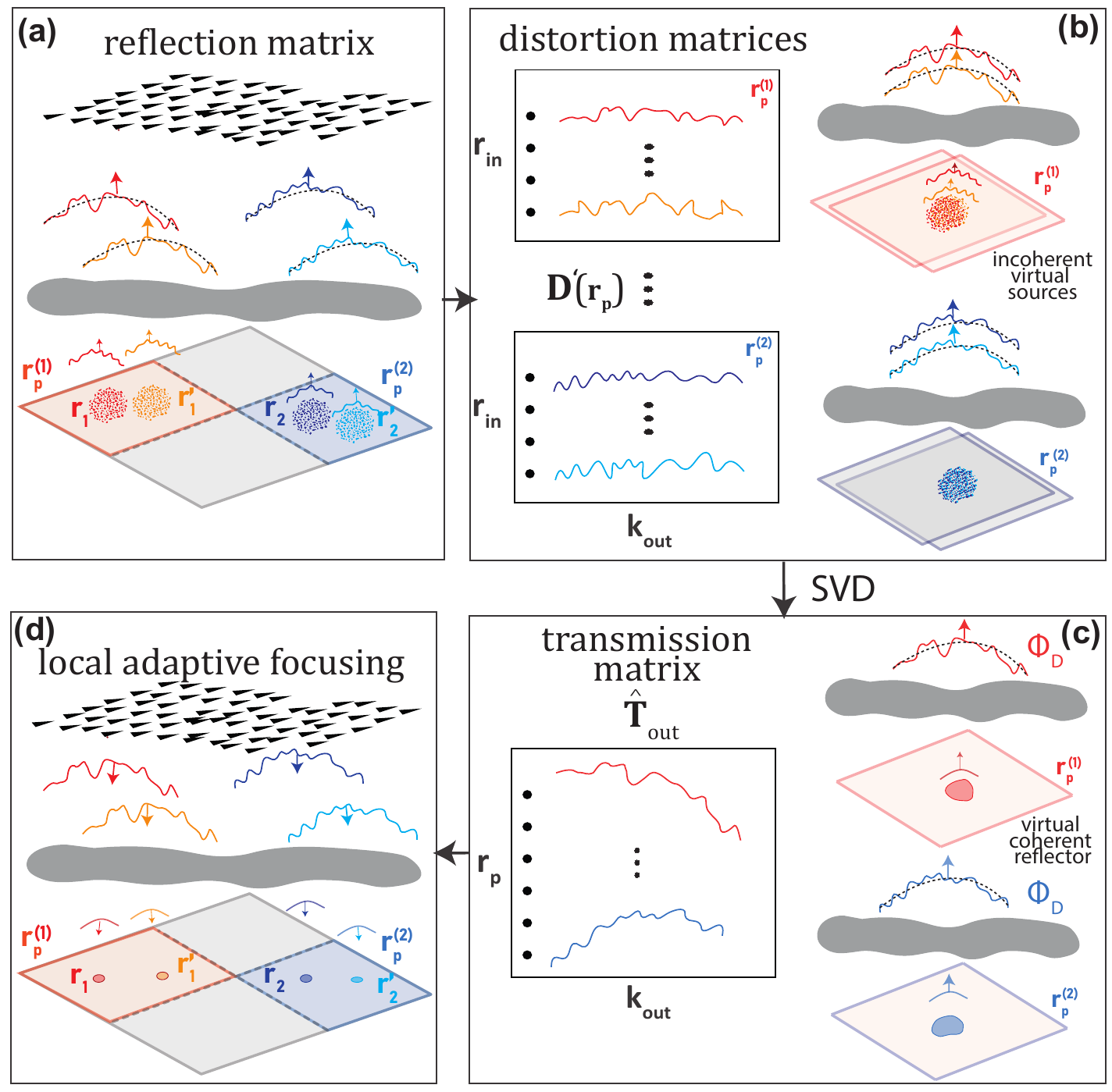}
 \caption{Local aberration correction. (a) One-side plane wave decomposition of each CLASS matrix yields the reflected wave-front associated with each focusing point $\mathbf{r}_\textrm{in}$. \rev{(b) By removing the geometrical curvature of each reflected wavefront (dashed line in a), one can study the phase distortions over each isoplanatic patch identified by their midpoint $\rp$. This operation amounts to realigning the wavefronts in each isoplanatic area as if they were generated by input focal spots virtually shifted to the origin. (c) SVD of each distortion matrix yields an aberration phase law $\phi_D$ for each spatial window by combining coherently each focal spot to synthesize a virtual coherent reflector. The set of aberration phase laws forms the estimator $\hat{\mathbf{T}}$ of the transmission matrix. (g) The phase conjugate of $\hat{\mathbf{T}}$ provides the focusing laws to compensate for phase distortions for each patch $\rp$.}} 
\label{chap3fig3}
 \end{figure*}

In this section, the distortion matrix concept is {applied to the CLASS FR matrix obtained in the previous section for compensation of spatially-distributed aberrations.} The process is outlined by five steps: $\textit{(i)}$ projection of the CLASS FR matrix at output into the plane wave basis (Fig.~\ref{chap3fig3}a), $\textit{(ii)}$ the realignment of the reflected wave-fronts to form a {distortion matrix $\mathbf{D}=[D(\mathbf{k}_\textrm{out},\bm{\rho}_\textrm{in},z)]$ (see Fig.~\ref{chap3fig3}b)}, $\textit{(iii)}$ the truncation of {$\mathbf{D}$} into local distortion matrices $\mathbf{D}'(\rp)$ , $\textit{(iv)}$ the singular value decomposition of {$\mathbf{D}'(\rp)$} to
extract a residual aberration phase law for each point $\rp$ and build an estimator \rev{$\hat{\mathbf{T}}$} of the transmission matrix $\mathbf{T}$ (Fig.~\ref{chap3fig3}\rev{c}); $\textit{(v)}$ {the phase conjugation of \rev{$\hat{\mathbf{T}}$}} to correct for output residual aberrations {(Fig.~\ref{chap3fig3}\rev{d})}. All of these steps are {then} repeated by exchanging output and input bases.



\subsection{The distortion matrix}

\rev{The output of the CLASS algorithm is a focused reflection matrix $\Rrr^{(C)}$ that still exhibits laterally-varying aberrations.
To assess these residual aberrations, the first step is to chose a basis in which the distortion of the CLASS wave-front is the most spatially-invariant. In a horizontally multi-layered medium such as NAFZ, the plane-wave basis is the most adequate since plane waves are the propagation invariants in this geometry. A plane-wave projection is performed at the output of the CLASS FR matrix {$\Rrr^{(C)}$}:
\begin{equation}
\label{Rkr}
    \mathbf{R}^{(C)}_{{\mathbf{k}\bm{\rho}}}(z)=\mathbf{P}^{'}_{0} \times {\Rrr^{(C)}}(z),
\end{equation}
$\mathbf{R}^{(C)}_{{\mathbf{k}\bm{\rho}}}(z)=[R^{(C)}(\kout,{\bm{\rho}_\textrm{in}},z)]$  connects each input focusing point $\rin{=(\bm{\rho}_\textrm{in},z)}$ to the CLASS wave-field in the plane wave basis (Fig.~\ref{chap3fig3}a).}

\rev{The CLASS wave-field can be understood as a sum of two components: (i) a geometrical component described by the reference matrix $\mathbf{P}^{'}_{0}$, containing the ideal wave-front generated by a source at $\rin$ {according to the propagation model described in Table~\ref{T1}} (dashed black curves in Fig.~\ref{chap3fig3}a ); (ii) a distorted component due {to spatially distributed aberrations that subsists after the CLASS procedure.} The latter component refers to the residual phase distortions that should be isolated from the CLASS wave-field in order to be properly compensated. This can be done by subtracting the ideal wave-front that would be obtained in absence of aberrations (i.e the geometrical component) from each CLASS wave-front induced by each input focusing wave at $\rin$.
Such operation can be expressed mathematically via a Hadamard product between $\mathbf{R}^{(C)}_{{\mathbf{k}\bm{\rho}}}(z)$ and $\mathbf{P}_0^{'}$. It yields the residual distortion matrix {$\mathbf{D}(z)$}:
\begin{equation}
{\mathbf{D}(z)  =  \mathbf{R}^{(C)}_{{\mathbf{k}\bm{\rho}}}(z)} \circ  \mathbf{P}_0^{'*} ,
\label{D_eq2}
\end{equation}
The matrix {$\mathbf{D}(z) $} connects any input virtual source $\rin$ to the residual distortion exhibited by the CLASS wave-field expressed in the plane wave basis (Fig.~\ref{chap3fig3}b). By removing the geometrical component of the CLASS wave-field, spatial correlations are highlighted between distorted wave-fields induced by neighbour virtual sources $\rin$~\citep{Badon2019}. Such correlations are a manifestation of a spatial invariance of residual aberrations over areas generally referred to as isoplanatic patches~\citep{lambert2021distor}.} 

\subsection{Local distortion matrices}

Our strategy is to divide the field-of-view into a set of overlapping regions (Fig.~\ref{chap3fig4}d). Each region is defined by a central midpoint {$\rp=(\bm{\rho}_p,z_p)$} and a spatial extension $L$. For each region, the local residual $\mathbf{D}$-matrix is defined as:
\begin{equation}
\label{Dkrp}
 {D'(\kout,\bm{\rho}_\textrm{in},\rp) =  D(\kout,\bm{\rho}_\textrm{in},z_p) W(\bm{\rho}_\textrm{in}-\bm{\rho}_p)},
\end{equation}
where {$W(\bm{\rho})$ is a spatial window function such that $W( \bm{\rho})=1$ for $|x| < L$ and $|y| < L$, and zero elsewhere}. Ideally, wave-front distortions should be invariant over each region, meaning that the virtual sources {$\rin=(\bm{\rho}_\textrm{in},z)$} associated with each region belong to the same isoplanatic patch. However, in practice, this hypothesis is not \revrita{fulfilled}. The isoplanatic length actually scales as the typical transverse dimension over which the wave velocity fluctuates. On the one hand, the dimension {$L$} of the window function should therefore be reduced to cover the smallest isoplanatic region as possible in order to provide a local and sharp measurement of aberrations. 
On the other hand, it should also be large enough to include a sufficient number of realizations of disorder in order to unscramble the effect of aberrations from the medium's reflectivity~\citep{lambert2021distor}. To reach a good estimate of the aberration phase law, the number of input focusing points in each region should be one order of magnitude larger than the number of resolution cells mapping the {CLASS focal spot (Fig.~\ref{chap3fig4}c)}~\citep{Lambert2020b}. 
{This is why the initial CLASS step was important to initiate the aberration correction process and reduce the extension of the focal spots before a local and finer compensation of residual aberration by means of the $\mathbf{D}$-matrix concept.} The area covered by the CLASS focal spot being {$20 \times 14$} $\textrm{km}^{2}$ (Fig.~\ref{chap3fig4}c), the extent of the window is chosen to be $55 \times 55$ $\textrm{km}^{2}$.

\subsection{Singular value decomposition}

\rev{Assuming \rev{local} isoplanicity {in each spatial window $W_L$} (Eq.~\ref{isoplanatic}), the coefficients of each distortion matrix $\mathbf{D}'(\rp) $ matrix can be expressed as follows~\citep{Lambert2020b}:
\begin{equation}
 D'(\mathbf{k}_\textrm{out}, \review{\bm{\rho}_\textrm{in}}, \rp ) = 
  {\delta \tilde{H}}_{\rev{L}}(\mathbf{k}_\textrm{out},\rp) \underbrace{ \int \review{d\bm{\rho} } \gamma(\review{\bm{\rho}+\bm{\rho}_\textrm{in}},z) \delta H_{\rev{L}}(\review{\bm{\rho}},\rp)  e^{i \mathbf{k}_\textrm{out}.\review{\bm{\rho}}}}_{\mbox{virtual source}} ,
  \label{Dcoef}
\end{equation}
This equation can be seen as a product between two terms: the output aberration transmittance and a virtual source term modulated by the medium's fluctuating reflectivity \review{$\gamma({\bm{\rho}},z)$}. The goal is now to unscramble these two terms in order to get a proper estimation of the aberration transmittance $\delta \tilde{H}_{\rev{L}}(\kout, \rp)$ at each point $\rp$.} 

\rev{In practice, this can be done through a {singular value decomposition (SVD)} of each local distortion matrix  {$\mathbf{D}'(\rp)$}:
\begin{equation}
\label{svdchp3}
{{\mathbf{D}'}(\rp)={\mathbf{U}}(\rp) \times \mathbf{\Sigma}(\rp) \times {\mathbf{V}}(\rp)^\dag}
\end{equation}
where $\Sigma$ is a diagonal matrix containing the real positive singular values $\sigma_i$ in a decreasing order $\sigma_1>\sigma_2> \cdots >\sigma_N$. {$\mathbf{U}(\rp)$ and $\mathbf{V}(\rp)$ are unitary matrices whose columns, 
$\mathbf{U}_i(\rp)=[U_i(\mathbf{k_\textrm{out}},\rp)]$ and $\mathbf{V}_i(\rp)=[V_i(\mathbf{r_\textrm{in}},\rp)]$}, correspond to the output and input singular vectors, respectively.}
\rev{The physical meaning of
this SVD can be intuitively understood by considering the asymptotic case of a point-like input focusing beam: $\delta H_{\rev{L}}(\bm{\rho},\rp)=\delta(\bm{\rho})$. In this ideal case, Eq.~\ref{Dcoef} becomes: ${D}'(\mathbf{k}_\textrm{out},\bm{\rho}_\textrm{in},\rp)= \delta \tilde{H}_{\rev{L}}(\mathbf{k}_\textrm{out},\rp) \gamma\left(\bm{\rho}_\textrm{in},z_p \right) $. Comparison with Eq.~\ref{svdchp3} shows that, a first approximation, $\mathbf{D}'(\rp)$ is of rank 1. The first output singular vector $\mathbf{U}_{1}(\rp)$ yields the residual aberration transmittance $\delta \tilde{\mathbf{H}}_{\rev{L}}(\rp) $ while the first input singular vector $\mathbf{V}_{1}(\rp)$ directly provides the medium reflectivity over the spatial window $W_L$. }

\rev{However, despite the CLASS correction, the input PSF $\delta H_L$ remains far from being point-like (Fig.~\ref{chap3fig4}c). The spectrum of $\mathbf{D}'(\rp)$ then displays a continuum of singular values but the first eigenstate of $\mathbf{D}'(\rp)$ is still of interest. $\mathbf{V}_{1}(\rp)$ corresponds to a rough estimate of the medium reflectivity that allows \revrita{realignment} in phase for each input focal spot. Therefore, the SVD process allows the synthesis of a coherent virtual reflector that can be leveraged for the estimation of the residual aberration transmittance $\delta \tilde{\mathbf{H}}_{\rev{L}}(\rp) $  (Fig.~\ref{chap3fig3}\rev{c}). More precisely, this is the normalized output singular vector $\mathbf{U}_{1}(\rp)=[U_1(\kout,\rp)/|U_1(\kout,\rp)|]$, that constitutes a relevant estimator for $\delta \tilde{\mathbf{H}}_{\rev{L}}(\rp) $~\citep{lambert2021distor}. The estimator of the transmission matrix is then given by \revrita{the Hadamard product}:}
\begin{equation}
\label{Ttransm}
     \rev{\hat{\mathbf{T}}}_\textrm{out}(\rp)= \mathbf{P}^{'}_{0} \circ e^{i [ \bm{\phi}_{D}^{(\textrm{out})}(\rp){+\phi_{C}^{(\textrm{out})} (z_p)}]}
\end{equation}
\rev{with $\bm{\phi}_{D}^{(\textrm{out})}(\rp)$, the phase of $\mathbf{U}_{1}(\rp)$}. The phase conjugate of $\rev{\hat{\mathbf{T}}}_\textrm{out}$ provides the focusing laws to compensate for the output phase distortions over each patch (Fig.~\ref{chap3fig3}\rev{d}). {The same method can be repeated by exchanging the focused and Fourier bases between input and output in order to estimate the transmission matrix $\mathbf{{T}_\textrm{in}}$~\citep{lambert2021distor}. The whole process is iterated once to refine the estimation of $\mathbf{{T}_\textrm{out}}$ and $\mathbf{{T}_\textrm{in}}$.}   

\subsection{Transmission matrix estimator}

The input phase laws obtained at the end of the aberration correction process are displayed in Fig.~\ref{chap3fig4}e for the central regions of the field-of-view highlighted in Fig.~\ref{chap3fig4}d. Although they show some similar features (in particular the low spatial frequency components), they also display some differences {that are} quantified by the correlation coefficient between the different phase masks with the central one. The value of this coefficient is reported below each phase mask. This correlation coefficient goes from $0.86$ for the closest spatial windows to $0.39$ for the furthest ones. {One can also notice that clear differences in the phase  \revrita{behaviour} can be observed between the north, center and the south of the field-of-view. The presence of these lateral differences is consistent with the three geological blocks in the region {(Fig.~\ref{chap3fig1}b)}.} The latter observation together with the correlation coefficient value show the importance of estimating a different phase law for each area and justifies the implementation of a local aberration correction process.

\subsection{Local compensation of spatially-distributed aberrations}

Using $\rev{\hat{\mathbf{T}}}_\textrm{out}$ and $\rev{\hat{\mathbf{T}}}_\textrm{in}$, a corrected FR matrix can be finally obtained:
\begin{equation}
\label{Rc}
    \mathbf{R}_{\mathbf{rr}}^{(D)}(z)=  \rev{\hat{\mathbf{T}}}^{\dag}_{\textrm{out}}   \times     \mathbf{R}_{\mathbf{kk}}(z) \times \rev{\hat{\mathbf{T}}}^{*}_{\textrm{in}} 
\end{equation}
The corresponding confocal image and RPSF are displayed at $z=25$ km in Figs.~\ref{chap3fig4}(d) and {(f)}. The comparison with their CLASS counterparts [Figs~\ref{chap3fig4}(a) and {(c)}] shows the \revrita{importance} of the local $\mathbf{D}-$matrix analysis. The diffuse background is clearly reduced and the RPSF is nearly similar to its ideal value (Fig.~\ref{chap3fig2}d), with almost all the backscattered energy contained in the white circle accounting for the diffraction limit. The residual background in Fig.~\ref{chap3fig4}{f} is probably associated with high-order aberrations whose coherence length (isoplanatic area) is smaller than the size $L$ of the window function $W(\bm{\rho})$.
\begin{table}[h!]
\centering
\begin{tabular} {|c|c|c|c|c|c|c|c| m{6.5cm} | m{2cm}| m{2cm}| m{2cm}| m{2cm}| m{2cm}| m{2cm}| m{2cm}| }
  \hline
  Correction steps & 0 & 1 & 2 & 3 & 4 & 5 & 6 \\ 
  \hline
    Correction type & 0& CLASS &  CLASS &  D &  D &  D &  D\\ 
  \hline
  Correction side & & Output & Input & Output & Input & Output & Input\\ 
  \hline
    Confocal gain (dB) & & 2.41 &  5.17 & 7.1 & 9 & 9.16 & 9.26 \\
  \hline
    Resolution $w$ (km) &40 & 20 &  8 & 7 & 6 & 6 & 6 \\
  \hline
\end{tabular}
\caption{{Confocal gain and  resolution at each step of the aberration correction process.}}
\label{T2}
\end{table}

To be more quantitative, a confocal gain can be computed from the intensity ratio between the corrected {(Fig.~\ref{chap3fig4}a and d)} and initial {(Fig.~\ref{chap3fig2}{e})} images. 
{The transverse resolution can also be estimated from the full width at half maximum $w$ of the RPSF. The confocal gain and the resolution are reported in Table~\ref{T2} at each step of the aberration correction process for depth $z=25$ km.} {{Strikingly, the transverse resolution is enhanced by a factor $\sim 7$ compared with its initial value and the confocal intensity is increased by more than 9 dB.}} {These values highlight the benefit of matrix imaging for in-depth probing of NAFZ at a large scale. }

In the next section, the 3D image of the medium around the NAFZ is {now revealed} by combining the images derived at each depth. A structural interpretation is then provided in light of previous studies on the NAFZ.

\section{3D structure of the NAFZ}
\begin{figure*}
 \centering
 \includegraphics[width=\textwidth]{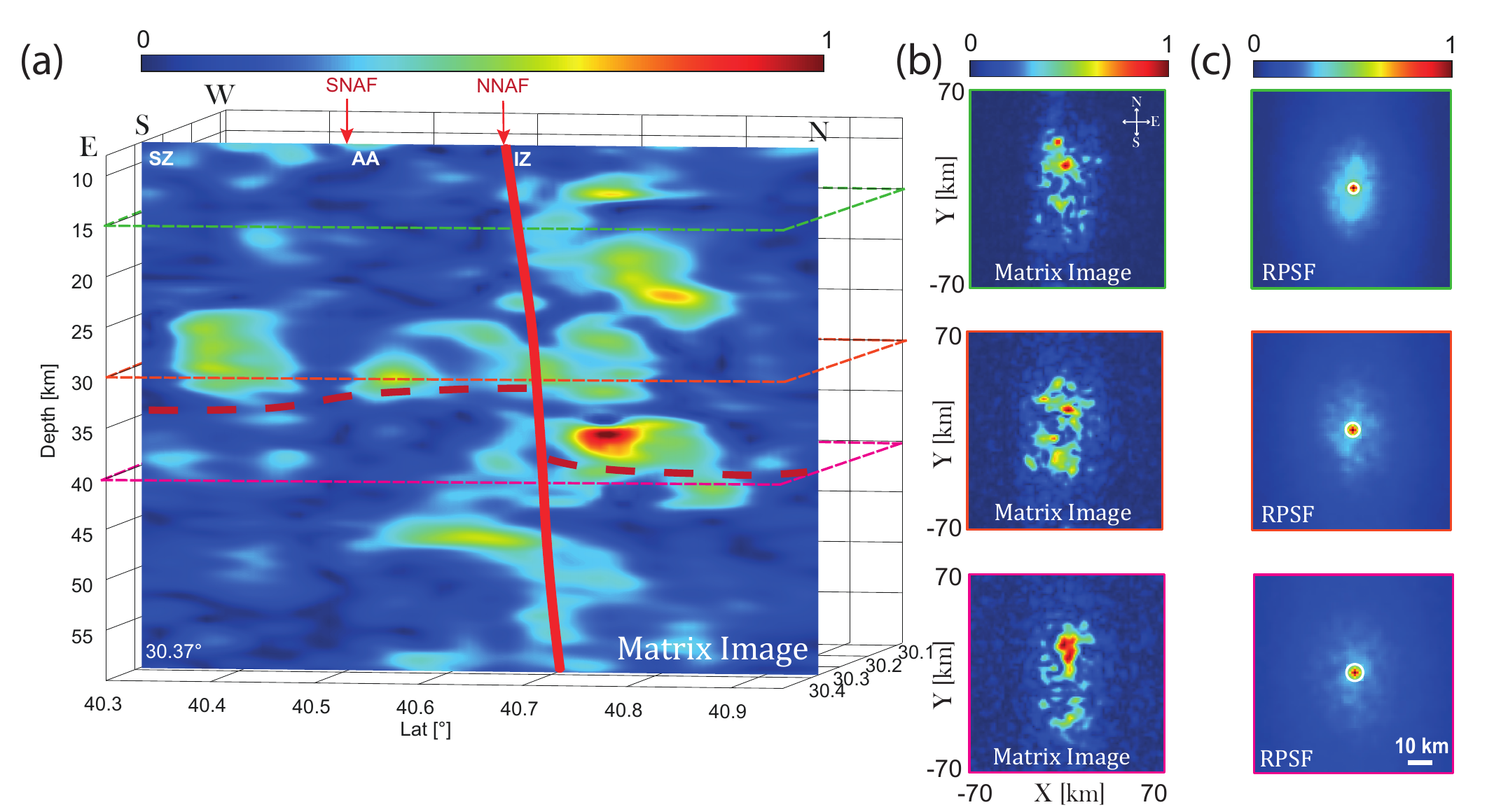}
 \caption{Final image of NAFZ. (a) Vertical North-South cross-section at $30.37\degree$E. The North-South profile is oriented perpendicular to the fault traces. The location of the profile is shown in Fig.~\ref{chap3fig1}a. The locations of the southern (SNAF) and northern (NNAF), and the major crustal blocks (SZ: Sakarya zone, AA: Armutlu-Almacik and IZ: Istanbul zone) are labeled.
 The interpreted location of the fault at depth are indicated by a red line.
 The color scale refers to the scattering intensity. It is normalized by the maximum value of the scattering energy inside the volume. 
Our interpretation of the Moho's location is indicated by red dashed lines. (b) Depth slices retrieved from the 3D scattering volume at $z=15, 30$ and $40$ km with (c) their corresponding {RPSFs}.}
\label{chap3fig5_2}
  \end{figure*}
{The previous sections have shown the process for a local compensation of phase distortions. Performing this correction process at each depth allows to uncover a well-resolved 3D image of the subsurface.} 

{\rev{Figure~\ref{chap3fig5_2}a} shows a North-South cross-section from the final 3D image. This cross-section is chosen at the same location as the one in Fig.~\ref{chap3fig5}\rev{a} and crosses the two fault strands. It also spans the three geological units: Istanbul zone (IZ), Armutlu-Almacik (AA) and Sakarya zone (SZ).
{The scattering generated by the heterogeneities of the medium induce a decrease of the backscattered energy with depth. Consequently, a drop of amplitude is observed in the 3D images. In order to compensate for this, the intensity in the cross-sections is normalized by the mean intensity calculated at each depth.} }

{Three depth slices retrieved from the final 3D images at $z=15, 30$ and $40$ km are also represented in \rev{Fig.~\ref{chap3fig5_2}b} with their corresponding RPSFs in \rev{Fig.~\ref{chap3fig5_2}c}. {Compared to the initial RPSFs (Figs.~\ref{chap3fig5}\rev{c}), {the resolution is significantly improved by a factor that \revrita{increases} from $7$ at small depth ($z<15$ km) to $9$ beyond $z=40$ km.}} The final matrix images in \rev{Fig.~\ref{chap3fig5_2}b} can also be compared to the initial confocal ones in Fig.~\ref{chap3fig5}\rev{b}. While the original images show random-like features, the images obtained after aberration correction reveal better defined features and a reflectivity that is mainly concentrated in the North. The same observation can be made by looking at the corrected cross-section in \rev{Fig.~\ref{chap3fig5_2}a}.}
The differences between the corrected and raw cross-sections are pronounced. While in the raw image, no clear structures and layers are visible, the corrected image reveals { sub-horizontal structures with a refined level of details, {thanks to the drastic gain in resolution revealed by the RPSF.}}

{Due to its significant seismic activity over the past 100 years, and to assess the ongoing hazard posed by this activity, extensive research has been conducted on the NAFZ to image its structure and determine its mechanical characteristics.
The scattering structure in \rev{Fig.~\ref{chap3fig5_2}a} is interpreted with reference to prior studies conducted in the region.}

{The first thing to notice in the profiles is that the scattered energy is predominantly situated in the North, which corresponds with the location of the Northern branch. This observation may be associated with the greater seismic activity of the Northern strand compared to the seismic activity of the Southern strand. 
The scattering below this strand and to the North of it, that extends to at least $60$ km, can be explained by the damage caused {by the large deformation of} this complex fault system {with a cumulative slip of the order of $80$ km~\citep{armijo1999,bohnhoff2016} during the last million years as well as the heterogeneities that have been inherited from the complex tectonic history of the region.} }

 {At the east of the Sea of Marmara, the Moho depth was reported to be between $30$ and $35$ km~\citep{zor2003,vanacore2013}. A deepening of the Moho was identified ($\sim 40$ km) in the IZ by~\cite{frederiksen},~\cite{taylor2016},~\cite{papaleo2017},~\cite{papaleo2018},~\cite{rost2021} and~\cite{jenkins2020}.
 In \rev{Fig.~\ref{chap3fig5_2}a}, a high scattering zone is observed between $25$ and $40$ km depth corresponding to a heterogeneous lower crust. Its lower boundary indicates the presence of the Moho (red dashed line). The Moho depth varies from $35$ km in the South to $42$ km in the North. The reflectivity is disrupted around $40.75\degree $N suggesting the presence of a step in the Moho below the Northern strand. The latter observation is in agreement with previous studies~\citep{rost2021,jenkins2020}.
Below the Moho, reflective structures are observed, mainly beneath AA and IZ, in agreement with~\cite{kahraman2015}.
These findings, supported with other studies~\citep{kahraman2015,papaleo2018,jenkins2020,rost2021}, suggest that the NNAF cuts though the entire crust and reaches the upper mantle~\citep{kahraman2015,papaleo2018,jenkins2020,rost2021}.} {A signature of the NAFZ in the mantle has also been proposed by the long period analysis of~\cite{fichtner2013}.}

{The signature of the Northern strand at depth can be identified by the presence of discontinuities in the scattering distribution in the first $20$ km of the crust (\rev{Fig.~\ref{chap3fig5_2}a}) and also by the termination of sub-Moho structures below the Northern strand. The Southern strand, on the other hand, lacks significant scattering, indicating that it has a weaker signal compared to the Northern strand. This, along with the continuity of the Moho in the South, suggests that the SNAF is confined in the crust and does not extend to the upper mantle, {Armutlu block being a crustal structure.}}

{In this section, only one cross-section has been depicted to demonstrate the significant enhancements and the gain in resolution provided by the presented matrix approach. A more in-depth analysis of the scattering volume around the NAFZ will be provided in a future study.}


\section{Conclusion}

{Matrix imaging} provides unprecedented view of the NAFZ. {To that aim, we exploited} seismic noise data from a dense deployment over the rupture region of the 1999 Izmit earthquake. {Ambient noise cross-correlations enable the passive measurement of the reflection matrix associated with the dense array of geophones. The body wave component is then used to image the in-depth reflectivity of the NAFZ subsurface.  
Compared with our previous work that considered a sparse scattering medium~\citep{touma2021}, the NAFZ case is more general since it exhibits both specular reflectors such as Moho discontinuity and a random distribution of heterogeneities.} 

{The strength of matrix imaging lies in the fact that it does not require an accurate velocity model. Here, a layered velocity model is employed but strong phase distortions subsist since lateral variations of the wave velocity are not taken into account. Nevertheless, such complex aberrations are compensated by two matrix methods previously developed in optical microscopy~\citep{kang2017,yoon2019} and ultrasound imaging~\citep{lambert2020,lambert2021distor}. First, the CLASS algorithm exploits angular correlations and memory effect exhibited by the reflection matrix to compensate for spatially-invariant aberrations. Second, a local analysis of the distortion matrix enables a local compensation of spatially-distributed aberrations. Together, those two approaches provide a sharp estimate of the transmission matrix between the Earth surface and the subsurface, leading to a narrowing of the imaging PSF by a factor {that goes from $7$ to $9$}. Therefore, a diffraction-limited resolution is reached for any pixel of the image. }

{Thanks to matrix imaging, the} scattering structure of the crust and upper mantle of the NAFZ continental strike slip fault {is thus} revealed. The $60$ km depth profile, show terminations of crustal discontinuities mainly below the northern branch. The localized scattering around the NNAF is consistent with the fact that it is the most seismically active fault and that it ruptured during the last $7.6$ Izmit earthquake. We identify a step in the Moho coinciding with the surface location of this branch in the East of DANA network. Moreover, the scattering extends to the upper mantle in the North. All these observations are consistent with previous studies and suggest that the NNAFZ is localized in the crust and extends to the upper mantle.


{\review{Even though the result are promising}, several points remain \revrita{that would allow improved images}. First, potential conversion between \review{S and P-waves} is not considered by matrix imaging. \review{The method could be improved in the future by considering both longitudinal and shear waves, as well as wave conversion between them.} Second, only a broadband compensation of phase distortions is performed. Yet, scattering phenomena or multiple reflections would require \revrita{a procedure that moves beyond} the application of simple time delays to the impulse response between geophones. Finally, a reflectivity image is only qualitative since it does not directly quantify the mechanical properties of the subsurface. Yet matrix imaging} {offers the possibility of} mapping the velocity distribution inside the medium~\citep{lambert2020}. This will be the focus of a future study.


\acknowledgments
This project has received funding from the European Research Council (ERC) under the European Union Horizon 2020 research and innovation program (grant agreement No 742335, F-IMAGE and grant agreement No. 819261, REMINISCENCE).

\noindent 
\textbf{Open Research}

\noindent \rev{The data used for this study were recorded by the temporary Dense Array for North Anatolia (DANA) and can be found at \citep{DANA2012}. The cross-correlation data and codes used to post-process the seismic data within this paper have been deposited at \citep{touma_leber_campillo_aubry_2023}.}

\vspace{\baselineskip}

\clearpage 

\clearpage

\renewcommand{\thetable}{S\arabic{table}}
\renewcommand{\thefigure}{S\arabic{figure}}
\renewcommand{\theequation}{S\arabic{equation}}
\renewcommand{\thesection}{S\arabic{section}}

\setcounter{equation}{0}
\setcounter{figure}{0}
\setcounter{section}{0}

\begin{center}
\Large{\bf{Supplementary Information}}
\end{center}
\normalsize

{This supplementary material includes further details about: {\alex{(i) the reflection point spread function;} (ii) {the} CLASS algorithm, (iii) the study of the reflection matrix \rev{in the plane wave basis} to determine the nature of the scattering process}.}

\section{Reflection point spread function}

\alex{{In this Supplementary Section, an analytical expression of the RPSF is derived} in the specular and scattering regimes. To do so, a local isoplanatic assumption is made (Eq.~17 of the accompanying paper). Under this assumption, the wave-field along each antidiagonal of the focused $\mathbf{R}-$matrix can be rewritten as follows (Eq.~12) 
\begin{equation}
\label{Rrr_coef2}
{{R}}({\bm{\rho}_m}-\Delta \bm{\rho}/2,{\bm{\rho}_m}+\Delta \bm{\rho}/2,z)= \int {d\bm{\rho}' } {H}_L(\bm{\rho}'-\Delta \bm{\rho}/2,\bm{\rho}_m,z ) \gamma({\bm{\rho}'+\bm{\rho}_m,z}) {H}_L(\bm{\rho}'+\Delta \bm{\rho}/2,\bm{\rho}_m,z).
\end{equation}
with $\bm{\rho}'=\bm{\rho}-{\bm{\rho}_m}$. Injecting Eq.~12 into Eq.~16 of the accompanying paper leads to the following expression for the RPSF:
\begin{align}
  I(\Delta {\bm{\rho}},\rev{\bm{\rho}}_m,z ) =  \int \int {d\bm{\rho}'_1 }d\bm{\rho}'_2 & {H}_L(\bm{\rho}'_1- \Delta \bm{\rho}/2,{\bm{\rho}}_m ,z ){H}_L^*(\bm{\rho}'_2- \Delta \bm{\rho}/2,{\bm{\rho}}_m,z ) \nonumber \\
  & \times \gamma({\bm{\rho}'_1+{\bm{\rho}}_\textrm{m},z})\gamma^*({\bm{\rho}'_2+{\bm{\rho}}_\textrm{m},z})  \nonumber \\
  & \times 
  {H}_L(\bm{\rho}'_1 + \Delta \bm{\rho}/2,{\bm{\rho}}_m,z) {H}_L^*(\bm{\rho}'_2+ \Delta \bm{\rho}/2,{\bm{\rho}}_m,z).
  \label{Ilocal}
\end{align}
To go beyond this general expression, one can consider two asymptotic regimes depending on the relative values between the characteristic length scale $\ell_\gamma$ of the reflectivity $\gamma(\bm{\rho},z)$ at the ballistic depth and the typical width $\delta \rho$ of the focal spots.}

\alex{In the specular scattering regime, the characteristic size
$\ell_\gamma$  of reflectors is larger than the width  $\delta \rho$ of each focal spot. $\gamma(\bm{\rho},z)$ can thus be assumed as invariant over the PSF support ($\gamma(\bm{\rho}_m+\bm{\rho}',z) \simeq \gamma(\bm{\rho}_m,z) $), such that:
\begin{align}
  I(\Delta {\bm{\rho}},\rev{\bm{\rho}}_m,z )& = |\gamma({{\bm{\rho}}_\textrm{m},z})|^2 \left | \int {d\bm{\rho}' } {H}_L(\bm{\rho}'- \Delta \bm{\rho}/2,{\bm{\rho}}_m ,z )
  {H}_L(\bm{\rho}' + \Delta \bm{\rho}/2,{\bm{\rho}}_m,z) \right |^2 \\
  &= |\gamma({{\bm{\rho}}_\textrm{m},z})|^2 \times  \left | {H}_L \stackrel{\Delta \bm{\rho}}{ \circledast} {H}_L (\Delta \bm{\rho},{\bm{\rho}}_\textrm{m},z) \right |^2
\end{align}
.}

\alex{In the scattering regime ($\ell_\gamma<\delta \rho$), the medium reflectivity can be considered, {in first approximation}, as random:}
\begin{equation}
\label{gamma_random}
\left\langle\gamma\left(\bm{\rho}_1\rev{,z}\right) \gamma^*\left(\bm{\rho}_2\rev{,z}\right)\right\rangle=\left\langle|\gamma|^2\right\rangle \delta\left(\bm{\rho}_2-\bm{\rho}_1\right),
\end{equation}
where $\delta$ is the Dirac distribution. \alex{The ensemble average of the RPSF (Eq.~\ref{Ilocal}) is given by:
\begin{align}
 \langle  I(\Delta {\bm{\rho}},\rev{\bm{\rho}}_m,z ) \rangle  =  \int \int {d\bm{\rho}'_1 }d\bm{\rho}'_2 & {H}_L(\bm{\rho}'_1- \Delta \bm{\rho}/2,{\bm{\rho}}_m ,z ){H}_L^*(\bm{\rho}'_2- \Delta \bm{\rho}/2,{\bm{\rho}}_m,z ) \nonumber \\
  & \times \left \langle \gamma({\bm{\rho}'_1+{\bm{\rho}}_\textrm{m},z})\gamma^*({\bm{\rho}'_2+{\bm{\rho}}_\textrm{m},z})  \right \rangle \nonumber \\
  & \times 
  {H}_L(\bm{\rho}'_1 + \Delta \bm{\rho}/2,{\bm{\rho}}_m,z) {H}_L^*(\bm{\rho}'_2+ \Delta \bm{\rho}/2,{\bm{\rho}}_m,z).
  \label{Ilocal2}
\end{align}
By combining the previous equation with Eq.~\ref{gamma_random}}, the ensemble average of $I(\Delta {\bm{\rho}},\bm{\rho}_m, z)$ \rev{can be expressed as follows:} 
\review{\begin{equation}
\rev{\left \langle I(\Delta {\bm{\rho}}, \bm{\rho}_m, z) \right \rangle} =\left\langle|\gamma|^2\right\rangle  \times \left[\left|H_{\rev{L}}\right|^2 \stackrel{\Delta {\bm{\rho}}}{\circledast}\left|H_{\rev{L}}\right|^2\right](\Delta {\bm{\rho}},\bm{\rho}_m, z) .
\end{equation}}

\section{CLASS {algorithm}}

{As stated in the accompanying paper, a {full-field} phase correction is {first} applied to the {$\Rkk$-}matrix through the CLASS algorithm~\citep{kang2017,choi2018}. }

\rev{In order to prove that $C(\kpl,z)$ is an estimator of the spatial frequency spectrum $\tilde{\gamma}\review{\left(\kpl , z\right)} $ of the medium reflectivity, one can inject the expression of $R\left(\kout , \kin , z\right)$ (Eq.~24 of the accompanying paper) into Eq.~26:
\begin{align}
    {C} \left(\kpl \rev{,z}  \right)= \tilde{\gamma}\review{\left(\kpl , z\right)} \sum_{\kout } \tilde{H}_{\rev{I}}\left(\kpl - \kout \rev{, z} \right) \tilde{H}_{\rev{I}}\left(\kout \rev{, z} \right).
\end{align}
${C} \left(\kpl \rev{,z}  \right)$ is thus equal to $\tilde{\gamma}\review{\left(\kpl , z\right)}$ modulated by the autocorrelation function of the aberration transmittance $\tilde{H}_I$. In first approximation, the latter quantity is real and the phase of ${C} \left(\kpl \rev{,z}  \right)$ is actually an estimator of the phase of $\tilde{\gamma}\review{\left(\kpl , z\right)}$.}

{In order to prove that $\phi_{C}(\k_{||} , \rev{ z} )$  is actually an estimator of $\mbox{arg} \left \lbrace \tilde{H}_{\rev{I}}(\k_{||} \rev{, z} ) \right \rbrace$ and to determine its bias, $\Rkk$ can be replaced by its expression (\rev{Eq. 24} of the accompanying paper) in \rev{Eqs.26-28 of the accompanying paper}. It yields the following expression for $\phi_{C}$: }
\begin{align}
\label{phioutExpressedWithH}
\phi_{{C}}\left(\kout \rev{, z} \right) =  &\arg \left[\tilde{H}_{\rev{L}}\left(\kout  \rev{, z} \right) \right] \nonumber\\
+ &\arg \left[\sum_{\kin} \lvert \tilde{\gamma}\left(\kout + \kin , z\right) \rvert^2 \tilde{H}_{\rev{L}}\left(\kin  \rev{, z} \right) \sum_{\kpr}  \tilde{H}_{\rev{L}}\left(\kpr, \rev{z} \right)  \tilde{H}^{*}_{\rev{L}}\left(\kout + \kin - \kpr \rev{, z}  \right)\right].
\end{align}
{The last expression shows that the estimator $\phi_C$ can be decomposed as a sum of its expectation $\arg \left[\tilde{H}_{\rev{L}}\left(\kout  \rev{, z}  \right) \right]$ and its bias. For a medium of random reflectivity, the term  $ \lvert \tilde{\gamma}\left(\kout + \kin , z\right) \rvert^2 $ can be replaced by its ensemble average, \textit{i.e} a constant. It yields:
\begin{align}
\label{phioutExpressedWithH2}
\phi_{C}\left(\kout  \rev{, z} \right) =  &\arg \left[\tilde{H}_{\rev{L}}\left(\kout , \rev{z}\right) \right] \nonumber\\
+ &\arg \left[  {H}_{\rev{L}}(\mathbf{0},\rev{z}) \sum_{\kpr}  \tilde{H}_{\rev{L}}\left(\kpr, \rev{z} \right)  \tilde{H}^{*}_{\rev{L}}\left(\kout + \kin - \kpr, \rev{z} \right)\right].
\end{align}
The last expression shows that the bias directly depends on the autocorrelation of the aberration phase law. The more complex the aberration is, the more biased its estimator is. It is equivalent to the bias exhibited by standard adaptive focusing methods induced by the blurring of a virtual guide star induced by focusing~\citep{lambert2021distor}. }

\rev{To prove that the CLASS operation amounts to maximize the confocal intensity in the focused basis, one can express the diagonal coefficients of $\Rrr^{(C)}$ as follows:
\begin{eqnarray}
R^{(C)}(\bm{\rho}_c,\bm{\rho}_c,z)&=&\sum_{\mathbf{k}_\textrm{out}}\sum_{\mathbf{k}_\textrm{in}}  R^{(C)}(\mathbf{k}_\textrm{out},\mathbf{k}_\textrm{in},z) e^{i (\mathbf{k}_\textrm{in}+\mathbf{k}_\textrm{out}) . \bm{\rho}_c} \nonumber \\
& = &\sum_{\kpl}  C'(\kpl) e^{i \kpl . \bm{\rho}_c}
\label{FT}
\end{eqnarray}
with ${C}' \left(\kpl \rev{,z}  \right)=\sum_{\kout } R^{(C)}\left(\kout, \kpl - \kout \rev{, z} \right)$, the sum of antidiagonal coefficients of $\Rkk^{(C)}$. The confocal image is thus the Fourier transform of $C'$. By realigning the phase of $\Rkk$'s coefficients located along the same antidiagonal, CLASS maximizes the intensity of ${C}'$ and thus of the confocal intensity, by virtue of Parseval identity: 
\begin{equation}
\sum_{\bm{\rho}_c} |R^{(C)}(\bm{\rho}_c,\bm{\rho}_c,z)|^2 \equiv \sum_{\kpl} |C'(\kpl)|^2.
\end{equation}
}

\section{Nature of the scattering process}

{The aberration correction process depends on the scattering regime we are facing. To determine it, the {plane wave} basis is particularly adequate~\citep{Lambert2020b}. This section shows how the $\Rkk$-matrix {(Eq. 14 of the accompanying paper)} can indicate the nature of the scattering processes taking place in NAFZ. }

\begin{figure}
 \centering
 \noindent  \includegraphics[width=14cm]{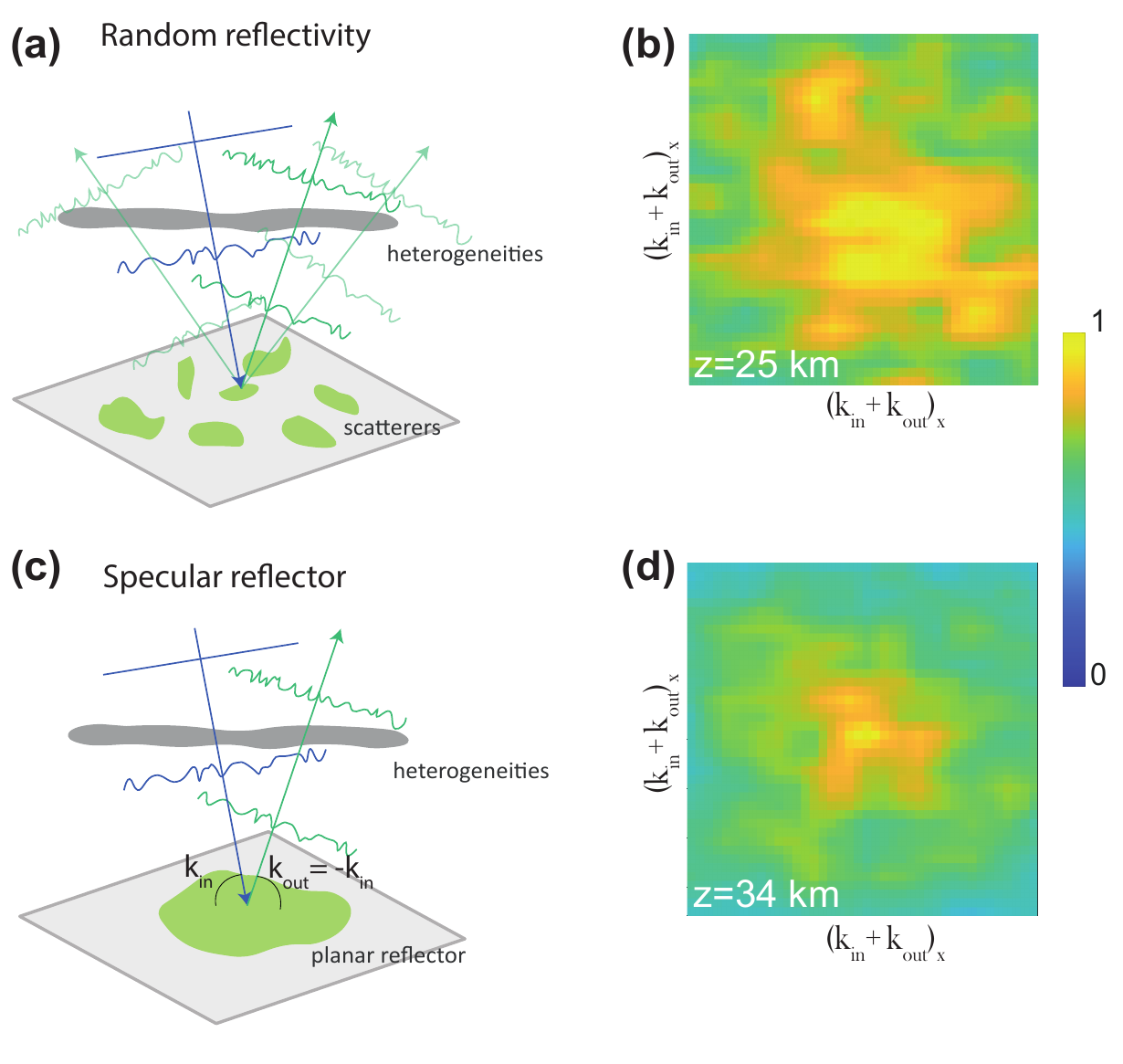}
 \caption{{Reflection matrix in the plane wave basis}. {(a) Sketch showing the {angular decomposition of the reflected wave-field in the speckle regime for a plane wave illumination (blue). A set of }plane waves (green) are reflected in all directions. {(b) Spatial} frequency spectrum of the reflectivity (Eq.~\ref{RkkModule}) at $z=25$ km. 
{(c)} Sketch showing the {angular decomposition of the wave-field reflected by a planar interface.} The incident plane wave (blue) is reflected with the same angle (green), such that {$\mathbf{k}_\textrm{out}+\mathbf{k}_\textrm{in}=\mathbf{0} $}; {(d) Spatial} frequency spectrum of the reflectivity (Eq.~\ref{RkkModule}) {at} $z=35$ km.}}
\label{chap3fig8}
  \end{figure}

{Indeed}, {assuming that the mismatch between the wave velocity model and reality only induces phase distortions between plane waves $ \left ( |\tilde{H}(\k)|{=1} \right )$,} the norm-square of {$\Rkk$-coefficients, $R(\mathbf{k}_\textrm{out},\mathbf{k}_\textrm{in}, z)$,} is shown to be independent of aberrations~\citep{Lambert2020b}:
\begin{equation}
\label{RkkModule}
{\left|R\left(\mathbf{k}_\textrm{out},\mathbf{k}_\textrm{in}, z\right)\right|^{2}=\left|\tilde{\gamma}\left(\mathbf{k}_\textrm{out}+\mathbf{k}_\textrm{in}, z\right)\right|^{2}.}
\end{equation}
{Each anti-diagonal of $\Rkk$ ({$\mathbf{k}_\textrm{out}+\mathbf{k}_\textrm{in}=$} constant) encodes one spatial frequency of the medium's reflectivity.
The spatial frequency spectrum of the medium's reflectivity can be estimated by averaging the intensity of the backscattered wave-field along each anti-diagonal of $\Rkk$.}
The result is displayed in Figs.~\ref{chap3fig8}{b and d} at two different depths: $z=25$ km and $z=35$ km. The norm square of the spatial frequency spectrum $\tilde{\gamma}(\k_{{||}},z)$ reveals the nature of the scattering process inside the medium. At $z=25$ km, the frequency spectrum shows an \rev{almost} flat spatial frequency spectrum (Fig.~\ref{chap3fig8}{(b)}) {which is a manifestation of \rev{distributed heterogeneities} {(Fig.~\ref{chap3fig8}(a))}. This regime is often referred to as a speckle wave-field in ultrasound imaging~\cite{Lambert2020b}. At depth $z=35$ km, $\tilde{\gamma}(\k_{{||}} , z)$ still shows a flat background due to randomly distributed heterogeneities but it also exhibits an over-intensity in the vicinity of {$\mathbf{k}_{||}=\mathbf{0}$ (Fig.~\ref{chap3fig8}(d))}. This peak in the spatial frequency domain is characteristic of a specular reflector (Fig.~\ref{chap3fig8}{(c)}) at that depth that may be associated with an interface between lower crust layers. From these two examples, we can see that the subsurface of NAFZ consists of a mix between specular reflectors and distributed heterogeneities.} 

\clearpage

\end{document}